\documentclass[nohyper]{JHEP3}
\usepackage{amsmath}
\usepackage{amsfonts}
\usepackage{slashed}
\usepackage{graphicx}
\usepackage{verbatim}
\usepackage{subfigure}
\usepackage{cite}

\newcommand \ip [2]{\langle #1|#2\rangle}

\title{General properties of multiparton webs: proofs from combinatorics}

\author{Einan Gardi\\
Scottish Universities Physics Alliance, and The Tait Institute,\\
School of Physics, The University of Edinburgh, Edinburgh EH9 3JZ, Scotland, UK\\
 E-mail: \email{Einan.Gardi@ed.ac.uk}}
\author{Chris D. White\\
Scottish Universities Physics Alliance, Department of Physics and Astronomy, University of Glasgow, Glasgow G12 8QQ, Scotland, UK\\
 E-mail: \email{c.white@physics.gla.ac.uk}}

\abstract{Recently, the diagrammatic description of soft-gluon exponentiation in scattering amplitudes has been generalized to the multiparton case. 
It was shown that the exponent of Wilson-line correlators is a sum of webs, where each web is formed through mixing between the kinematic factors and colour factors of a closed set of diagrams which are mutually related by permuting the gluon attachments to the Wilson lines. 
In this paper we use replica trick methods, as well as results from enumerative combinatorics, to prove that web mixing matrices are always: (a) \hbox{\emph{idempotent}}, thus acting as projection operators; and (b) have \emph{zero sum rows}: the elements in each row in these matrices sum up to zero, thus removing components that are symmetric under permutation of gluon attachments. 
Furthermore, in webs containing both planar and non-planar diagrams we 
show that the zero sum property holds separately for these two sets.
The properties we establish here are completely general and form an important step in elucidating the structure of exponentiation in non-Abelian gauge theories. 
}


\preprint{Edinburgh 2011/04}
\begin{document}

%%%%%%%%%%%%%%%%%%%%%%%%%%%%%%%%%%%%%%%%%
\section{Introduction}
\label{introduction}

Correlators of Wilson lines are important in many applications of gauge field theories in both the perturbative and non-perturbative regimes. 
Recently Wilson lines have been playing a major role in exploring the properties of scattering amplitudes in non-Abelian gauge theories. The best known example is the conjectured duality in ${\cal N}=4$ supersymmetric Yang-Mills theory at large $N_c$, between scattering amplitudes and the vacuum expectation value of polygon Wilson-loops in an auxiliary coordinate space. This relation was first proposed by Alday and Maldacena~\cite{Alday:2007hr} to hold at strong coupling, where the
AdS/CFT correspondence~\cite{Maldacena:1997re} gives a handle on the computation. Immediately thereafter it was discovered that the relation holds also at weak coupling~\cite{Drummond:2007aua,Brandhuber:2007yx}. This progress, hinting at integrable structures, 
stimulated much further 
work, e.g.~\cite{Alday:2008yw,Alday:2008zza,Drummond:2008aq,Drummond:2007bm,Drummond:2007au,Drummond:2007cf,Brandhuber:2007yx,DelDuca:2010zp,DelDuca:2010zg,
DelDuca:2009au,Heslop:2010kq,Brandhuber:2010bj,Heslop:2010xa,
Brandhuber:2009da}, leading to better understanding of the symmetries that dictate the structure of the amplitude, and further remarkable relations~\cite{Eden:2010ce,Mason:2009qx,ArkaniHamed:2009dn,ArkaniHamed:2009si}. For a recent review see \cite{Beisert:2010jr,Drummond:2010km,Alday:2010kn}.

Another important example is given by correlators of semi-infinite Wilson-line rays branching out of a local interaction vertex, where an arbitrary colour exchange occurs. Such correlators provide an effective-theory description of soft gluon interactions with energetic partons participating in a hard scattering process. Each hard parton in the scattering amplitude is replaced, in the ``eikonal'' approximation, by a Wilson line along its classical trajectory, providing a source for the soft-gluon field. The recoil of the hard parton due to the interaction with the soft gluons is neglected. 
Owing to the factorization of soft modes with respect to hard and collinear ones~\cite{Mueller:1979ih,Sen:1981sd,Sen:1982bt,Collins:1980ih,Magnea:1990zb,Kidonakis:1998nf,Sterman:2002qn}, many properties of the scattering amplitude, notably its infrared singularity structure, are captured by this description. Importantly, this picture is valid for general $N_c$, where there is interesting interplay between colour-flow and kinematics.
Because Wilson line correlators are, in general, much simpler than the corresponding amplitudes, this effective description is of prime importance for studying scattering amplitudes.
The Wilson-line description allows access to all-order properties in perturbation theory, as well as to strong coupling limit methods.

The fundamental property of an operator made of Wilson lines is that it renormalizes multiplicatively~\cite{Polyakov:1980ca,Arefeva:1980zd,Dotsenko:1979wb,Brandt:1981kf} (see also ~\cite{Korchemsky:1985xj,Ivanov:1985np,Korchemsky:1987wg,Korchemsky:1988hd,Korchemsky:1988si}). Consequently Wilson-line correlators exponentiate, 
\begin{equation}
{\cal S}=  \,\,\,{\cal P} \,\exp\bigg\{-\frac12 \int_0^{\mu^2} \frac{d\lambda^2}{\lambda^2} \Gamma_{\cal S}(\lambda^2)\bigg\}\,,
\label{Ssol}
\end{equation}
and their structure is encoded, to all orders, in the ``soft anomalous dimension'' matrix $\Gamma_{\cal S}$, which is itself a matrix in colour-flow space (hence the ordering operator ${\cal P}$), encoding both colour and kinematic dependence. 

The analysis of the soft anomalous dimension has been the basis of much theoretical work in recent years leading to substantial progress in understanding the structure of infrared (long-distance) singularities in multi-leg amplitudes, developments that are important both from the field-theoretic perspective and the collider-physics one.  Infrared singularities of scattering amplitudes have been fully determined to two-loop order, with both massless~\cite{Aybat:2006wq,Aybat:2006mz} and massive partons~\cite{Kidonakis:2009ev,Mitov:2009sv,Becher:2009kw,Beneke:2009rj,Czakon:2009zw,Ferroglia:2009ep,Ferroglia:2009ii,Kidonakis:2009zc,Chiu:2009mg,Mitov:2010xw,Ferroglia:2010mi}.
Moreover, in the massless case, stringent all-order constraints were 
derived~\cite{Becher:2009cu,Gardi:2009qi,Becher:2009qa} based on factorization and rescaling symmetry, leading to a remarkable possibility, 
namely that all soft singularities in any multi-leg amplitude take the form of a sum over colour dipoles formed by any pair of hard coloured partons.
Despite recent progress~\cite{Dixon:2008gr,Gardi:2009qi,Dixon:2009gx,Becher:2009cu,Becher:2009qa,Dixon:2009ur,Gardi:2009zv,Gehrmann:2010ue}, the basic questions of whether the sum-over-dipoles formula receives corrections, and at what loop order, remain so far unanswered. Further progress in understanding the singularity structure of multiparton scattering amplitudes in both the massless and massive cases requires new techniques to facilitate higher-loop computations. 

An alternative approach to non-Abelian exponentiation, aiming at a direct diagrammatic construction of the exponent, is that of `webs'~\cite{Sterman:1981jc,Gatheral:1983cz,Frenkel:1984pz,Gardi:2010rn,Mitov:2010rp}. 
In an Abelian theory, webs -- the diagrams that contribute to the exponent -- are simply the set of all connected diagrams\footnote{The term `connected diagrams' excludes the eikonal lines themselves: diagrams with gluons (or photons) that are attached to the same eikonal line(s) are not considered connected. Examples can be found in Section~1 of \cite{Gardi:2010rn}.}, since disconnected diagrams 
would be generated by expanding the exponential.
In a non-Abelian theory complications arise due to the fact that multiple gluon attachments to a given Wilson line give rise to a sequence of non-commuting colour generators. Nevertheless, it has long been known~\cite{Sterman:1981jc,Gatheral:1983cz,Frenkel:1984pz} that for a Wilson loop -- or two eikonal lines meeting at a cusp, where a \emph{colour singlet} hard interaction takes place -- the concept of webs naturally generalises to the non-Abelian theory: the expression for a Wilson loop is then
\begin{equation}
\label{web_exponentiation}
{\cal S}=  \,\, \,\exp \bigg\{\sum_{D} {\cal F}(D)\,\widetilde{C}(D)\,\bigg\}\,,
\end{equation}
where for any given diagram $D$, ${\cal F}(D)$ and $\widetilde{C}(D)$ denote, respectively, the kinematic dependence and the ``Exponentiated Colour Factor'' (ECF), where the latter is distinct from the conventional colour factor of the diagram, $C(D)$. This replacement encapsulates the non-Abelian nature of the interaction.
Furthermore, the diagrams that contribute to the exponent can be characterized as those which cannot be partitioned by cutting only the Wilson lines\footnote{A useful review of these concepts, followed by treatment of the three eikonal line case can be found in Chapter 3 of \cite{Berger:2003zh}.}.  These diagrams are sometimes referred to as ``two-eikonal-line irreducible'' or ``colour connected''.  

Recently, the concept of webs has been further generalized to address non-Abelian exponentiation in the multiparton case~\cite{Gardi:2010rn,Mitov:2010rp}. Considering $L$ Wilson lines branching out of a local interaction vertex, where some arbitrary colour exchange occurs, the two groups of authors have shown that exponentiation in the form of eq.~(\ref{web_exponentiation}) survives, although the simple topological criterion of irreducibility does not. Ref.~\cite{Gardi:2010rn} provided an explicit formula\footnote{Both papers provided an algorithmic way to determine the ECF $\widetilde{C}(D)$ using an inverse relation where conventional colour factors are written as a linear combination of ECF's. We show that the two are equivalent in appendix~\ref{sec:inverse_formulae_comparison}.} for the ECF's $\widetilde{C}(D)$ in terms of conventional colour factors.
Diagrams at arbitrary loop order were found to form closed sets
containing diagrams related by permutations of gluon attachments to the external Wilson lines. The ECF of a given diagram $D$ is a linear combination of conventional colour factors of diagrams $D'$ belonging to the same closed set, namely
\begin{equation}
\widetilde{C}(D)=\sum_{D'}R_{DD'}C(D'),
\label{Ctilde}
\end{equation}
where $R_{DD'}$ is a \emph{web mixing matrix}. 

The emerging generalization of a web in the multiparton case is therefore the entire set of diagrams
whose colour factors mix: those which are mutually related by permuting the gluon attachments to the Wilson lines. Each such set of diagrams can be labelled by the number of gluon attachments~$n_k$ to each of the Wilson lines $k=1\ldots L$. Distinct diagrams $D$ in the set differ only by the order of attachments of the gluons to each line. 
The contribution of each web to the exponent is
\begin{equation}
W_{(n_1,n_2,\ldots,n_L)}\,\equiv\, \sum_{D}{\cal F}(D)\,\, \widetilde{C}(D)\,,
\label{exp1}
\end{equation}
where, as above, ${\cal F}(D)$ is the kinematic part of diagram $D$ and $\widetilde{C}(D)$ is its ECF. 
Substituting 
eq.~(\ref{Ctilde}), we may rewrite this as a double sum
\begin{equation}
W_{(n_1,n_2,\ldots,n_L)}\,=\,\sum_{D,D'}{\cal F}(D)\,\,R_{DD'}\,\,C(D')\,= {\cal F}^T R C\,,
\label{exp2}
\end{equation}
which makes the role of the web mixing matrix explicit. As is clear from 
eq.~(\ref{exp2}), we may think of this matrix as acting either on the vector of
conventional colour factors for each web, or as acting on the (transposed) vector of
kinematic parts. The web mixing matrices thus encode subtle relationships
between the kinematic and colour structure of the exponent, and the further 
study of these matrices is crucial to understanding the all-order structure of scattering amplitudes.

Having at hand an explicit formula for determining the ECF's, Ref.~\cite{Gardi:2010rn} examined several classes of diagrams, exploring the properties of the web mixing matrices and their effect on the singularity structure of the exponent. Through these examples 
a couple of interesting properties were noted~\cite{Gardi:2010rn}:
\begin{enumerate}
\item {\it Idempotence}: for any web mixing matrix $R$, one has $R^2=R$, or
\begin{equation}
R_{DE}=\sum_{D'}R_{DD'}R_{D'E}\,\qquad\quad \forall D,E\,.
\label{idemp}
\end{equation}
As a consequence, any mixing matrix $R$ is diagonalizable, and its
eigenvalues can only be 0 or~1. Of course, both eigenvalues will generically be degenerate.
As explained in~\cite{Gardi:2010rn}, $R$ can therefore be interpreted as a genuine projection operator, which selects, as the eigenvectors corresponding to eigenvalue $1$, those combinations of colour factors and kinematic factors that build up the exponent. In contrast the eigenvectors corresponding to eigenvalue $0$ do not enter the exponent: they correspond precisely to those contributions that are generated from lower order webs by expanding the exponential. Ref.~\cite{Gardi:2010rn} further demonstrated that this structure is intimately related to the cancellation of subdivergences\footnote{Here we only discuss unrenormalized webs. Upon renormalizing  the multieikonal vertex, additional cancellations take place involving commutators of lower order webs and counter-terms~\cite{Mitov:2010rp}.} in the exponent. 
\item {\it Zero sum rows}: for any mixing matrix $R$, the elements in any row sum to zero,
\begin{equation}
\sum_{D'}R_{DD'}=0\,\qquad\quad \forall D.
\label{zerosum}
\end{equation}
This amounts to the fact that terms that are
fully symmetric in colour under permutation of the attachments to the Wilson lines do not contribute to the exponent, but rather are generated by the exponentiation of lower-order webs~\cite{Gardi:2010rn}. 
In other words, this is the generalization of the 
``maximally non-Abelian'' nature of webs~\cite{Gatheral:1983cz} from the two parton to the multiparton case.
\end{enumerate}
In~\cite{Gardi:2010rn} these properties were conjectured to hold for any web, based on explicit examples up to four loop order. The aim of the present paper is to \emph{prove} that these properties indeed hold in general.

In order to prove the above conjectures, it will be useful to recall
techniques that were already used in~\cite{Laenen:2008gt,Gardi:2010rn} 
to establish the structure described above and compute the web mixing matrices. 
We will use the replica trick, a technique borrowed from statistical physics (see 
e.g.~\cite{Replica}) which shortcuts the combinatorics involved in deriving
exponentiation properties. In particular, we will see that the replica
trick provides an elegant explanation of why idempotence must be an inherent property of web mixing matrices. 
The proof of the zero sum row property relies
upon a closed form combinatoric formula for ECF's given 
in~\cite{Gardi:2010rn}, which is further developed here. We show that it is possible to relate this formula to known results from the theory of integer partitions, and the zero sum row
property then emerges from known combinatoric identities.

The structure of the paper is as follows. In section~\ref{sec:replica}
we review the replica trick formalism of Ref.~\cite{Gardi:2010rn}, and then,
in section~\ref{sec:idemp}, use it to prove the idempotence property. In section \ref{sec:comb} we recall the formula for ECF's derived in Ref.~\cite{Gardi:2010rn}, and introduce the concept of `overlap functions' in order to derive an explicit expression for the mixing matrices; the latter is then used in section~\ref{sec:zerosum} to prove the zero sum property. In section~\ref{sec:planar} 
we discuss some further constraints on the mixing matrices based on the planar limit.
Finally, in section~\ref{sec:conclude} we conclude with a short discussion of our results.
We also include two appendices: In appendix~\ref{sec:inverse_formulae_comparison} we explain the equivalence between the combinatoric formulae presented in~\cite{Gardi:2010rn} and those obtained using the alternative approach of~\cite{Mitov:2010rp}. Appendix \ref{app:stirling} summarizes useful properties of the Stirling numbers of the second kind, which emerge in the proof of the zero sum property.

%%%%%%%%%%%%%%%%%%%%%%%%%%%%%%%%%%%%%%%%%
\section{The replica trick}
\label{sec:replica}

In this section, we recall the replica-trick formalism that was used 
in~\cite{Gardi:2010rn} to establish the existence of web mixing matrices and compute them. In the next section we will use a similar method to prove that web mixing matrices are 
idempotent. 

Our starting point is to consider a hard interaction 
$H(x_1,\ldots x_L)_{a_1\ldots a_L}$ which produces 
$L$ coloured particles (partons) with colour indices $a_k$ at 
4-positions~$x_k$. The scattering amplitude for such an interaction, dressed by any
number of soft (eikonal) gluon emissions, may be 
written in the path integral representation~\cite{Laenen:2008gt,Gardi:2010rn}:
\begin{equation}
{\cal M}_{b_1\ldots b_L}(p_1,\ldots,p_L)=\int \left[  {\cal D}{A}^\mu_s \right] H_{a_1\ldots a_L}(0,\ldots, 0)\, {\rm e}^{{\mathrm i}S[A^\mu_s]}\prod_k
\left({\cal P}\exp\left[{\mathrm i}g_s\int dt\beta_k\cdot{A}_s\right]\right)_{a_kb_k}\,,
\label{amppath}
\end{equation}
where $A_s^\mu$ is the soft gauge field with action $S[A_s^\mu]$.
Associated with each external line is a
Wilson line factor describing soft gluon emissions, where the trajectory
of the $k^{\rm th}$ particle having 4-velocity $\beta_k$, is a straight line, 
$z_k(t)=x_k+t\beta_k$.
We have used the fact that, at eikonal level, one may set $x_k=0$ in 
eq.~(\ref{amppath})~\cite{Laenen:2008gt}, making the hard interaction effectively local and independent of $A_s^\mu$. Taking the hard interaction outside the path integral, one may write
\begin{equation}
{\cal M}_{b_1\ldots b_L}(p_1,\ldots,p_L)=H_{a_1\ldots a_L}{\cal Z}_{a_1\ldots a_L,b_1\ldots b_L},
\label{amppath2}
\end{equation}
which makes explicit the fact that soft gluons are described by the matrix
\begin{equation}
{\cal Z}_{a_1\ldots a_L,b_1\ldots b_L}=\int\left[ {\cal D}{A}^\mu_s\right]\,{\rm e}^{{\mathrm i}S[A^\mu_s]}\prod_k
\left({\cal P}\exp\left[{\mathrm i}g_s\int dt\,\beta_k\cdot{A}_s\right]\right)_{a_kb_k}.
\label{Zdef}
\end{equation}
This has the form of a generating functional for a quantum field theory
for the soft gauge field. The Wilson line factors act as source terms
coupling the gauge field to the outgoing hard partons. Feynman diagrams in this
theory are {\it subdiagrams} in the full theory, which span the external
parton lines. %(e.g. figure~\ref{abex1}(c)). 
Thus, exponentiation of diagrams
in the quantum field theory of eq.~(\ref{Zdef}) amounts to the exponentiation
of soft gluon subdiagrams in the full theory. 
In this paper we shall refer to them as diagrams -- not subdiagrams -- as we directly consider here correlators of Wilson lines, not partonic amplitudes.
By definition, diagrams which exponentiate will be referred to as webs.
We have restricted ourselves to the eikonal approximation in 
eq.~(\ref{amppath}) so as to simplify equations in what follows. However,
as explained in~\cite{Gardi:2010rn}, conclusions reached about the 
exponentiation of eikonal corrections can also be extended to next-to-eikonal
order (see~\cite{Laenen:2010uz} and Refs. therein). 

Projecting the amplitude of eq.~(\ref{amppath2}) onto a basis of colour
tensors corresponding to distinct colour-flows (see~\cite{Kidonakis:1998nf}), we have
\[
{\cal M} ={\cal M}_J \, c^J_{b_1\ldots b_L}\,,
\]
where summation over the colour-flow index $J$ is understood.
One may then rewrite eq.~(\ref{amppath2}) for each component in colour-flow space,
\begin{equation}
{\cal M}_J=\sum_{I}H_I{\cal Z}_{IJ},
\label{amppath3}
\end{equation}
where 
\begin{equation}
{\cal Z}_{IJ}=\int[{\cal D}A_s^\mu]\,{\rm e}^{{\mathrm i}S[A_s^\mu]}\left[
\Phi_1\otimes\ldots\otimes\Phi_L\right]_{IJ}
\label{Zdef2}
\end{equation}
is the soft gluon generating functional, and where we denote the Wilson line factor associated with the $k^{\rm th}$ parton line by 
\begin{equation}
\label{Phi_k}
\Phi_k={\cal P}\exp\left[{\mathrm i}g_s\int dt\,\beta_k\cdot{A}_s\right].
\end{equation}
The $\otimes$ symbol is used in (\ref{Zdef2}) ($L-1$ times) since each Wilson line 
carries distinct partonic colour indices in some representation. 
Equation~(\ref{Zdef2}) makes clear that the soft gluon generating functional
is matrix-valued in colour flow space. This imbues the quantum field theory
for the soft gauge field with a non-trivial degree of combinatorics, so that 
further work is involved in ascertaining which diagrams exponentiate in the
theory, and what their associated colour factors are. 
>From now on we suppress the colour flow indices $I$ and $J$ for brevity.

One may derive the exponent of the generating functional using the replica
trick~\cite{Replica}\footnote{For other 
applications of the replica trick in high energy physics, 
see~\cite{Arefeva:1983sv,Fujita:2008rs,Akemann:2000df,
Damgaard:2000gh,Damgaard:2000di}.}, as explained in~\cite{Gardi:2010rn}. The argument proceeds as 
follows. Firstly, one considers a {\it replicated theory} consisting of
$N$ identical copies of the soft gauge
field, each with the same action and source terms as in eq.~(\ref{Zdef2}). 
That is, each gauge field has its usual self-interactions, but fields in a given replica do not
interact with ones in other replicas. The generating function for the replicated
theory is given by
\begin{equation}
{\cal Z}^N=\int[{\cal D}A_1^\mu]\ldots[{\cal D}A_N^\mu]
{\rm e}^{{\mathrm i}\sum_iS[A_i^\mu]}\left[
(\Phi_1^{(1)}\ldots\Phi_1^{(N)})\otimes\ldots\otimes
(\Phi_L^{(1)}\ldots\Phi_L^{(N)})\right].
\label{Zrepdef}
\end{equation}
Here we have dropped the subscript {\it s} to denote the soft gauge field,
and used $A_i$ to represent a gauge field with replica number $i$ (with
$N$ the total number of replicas). The action for the replicated theory is
given by the sum of the individual replica actions, due to the fact that
the replicas do not interact with each other. Finally, $\Phi_k^{(i)}$ is a
Wilson line factor associated with replica number $i$ and parton line
$k$: it provides a source for soft gluons of replica $i$, $A_i$, in the colour 
representation corresponding to parton $k$. 

We have recognized on the left-hand side of eq.~(\ref{Zrepdef})
that the generating functional for the replicated theory is related to
the original generating functional raised to the power $N$, a direct
consequence of the fact that the replicas are non-interacting.
Each parton line now carries a product of $N$ Wilson line factors, 
ordered along each parton line (away from the hard interaction vertex) 
in terms of increasing replica number. This product has the form\footnote{We have not written explicitly the indices on individual path-ordered exponentials. It is essential, however, that these are matrix-valued, so their order is important.}
\begin{equation}
\Phi_k^{(1)}\ldots\Phi_k^{(N)}=\left({\cal P}\exp\left[
\int dt\beta_k\cdot A_1\right]\right)\ldots\left({\cal P}\exp\left[
\int dt\beta_k\cdot A_N\right]\right).
\label{wilsonprod1}
\end{equation}
One cannot immediately read off the Feynman rules from a product of
path-ordered exponentials. Instead, one needs to rewrite 
eq.~(\ref{wilsonprod1}) as a single path-ordered exponential. One may do 
this by first noting that upon expanding the product of exponentials in
eq.~(\ref{wilsonprod1}), each term contains a product of gauge field
operators, ordered according to increasing replica number. Thus one may
write~\cite{Gardi:2010rn}
\begin{equation}
\left({\cal P}\exp\left[
\int dt\beta_k\cdot A_1\right]\right)\ldots\left({\cal P}\exp\left[
\int dt\beta_k\cdot A_N\right]\right)=
{\cal R}{\cal P}\exp\left[\sum_i\int dt\beta_k\cdot A_i\right],
\label{wilsonprod2}
\end{equation}
where ${\cal R}$ is a {\it replica-ordering operator} which reorders any
product of gauge fields to ensure that the replica numbers are increasing. 
The generating functional for the replicated theory, eq.~(\ref{Zrepdef}),
may thus be rewritten as
\begin{align}
{\cal Z}^N&=\int[{\cal D}A_1^\mu]\ldots[{\cal D}A_N^\mu]\,\,
{\rm e}^{{\rm i}\sum_{i=1}^N S[A_i^\mu]}\,\, \times\notag\\&\hspace*{60pt}
{\cal R}\left\{{\cal P}\exp\left[\sum_{i=1}^N\int dt
\beta_1\cdot A_i\right]\otimes
%\right.\notag\\&\left.\quad
\ldots\otimes {\cal P}\exp\left[\sum_{i=1}^N\int dt
\beta_L\cdot A_i\right]\right\}.
\label{Zrepdef2}
\end{align}
Feynman diagrams in this replicated theory have kinematic parts which are
the same as the topologically similar diagrams in the original theory. 
However, the colour factors are different in the replicated theory due to 
the presence of the ${\cal R}$ operator. That is, colour matrices are reordered
on each parton line so as to satisfy the replica ordering constraint. 

The reason for replicating the original theory is as follows~\cite{Gardi:2010rn}. Upon expanding in powers of $N$,
\begin{equation}
{\cal Z}^N=1+N\log{\cal Z}+{\cal O}(N^2)\,.
\label{ZpowN}
\end{equation}
Applying such an expansion to (\ref{Zrepdef2}), and picking the ${\cal O}(N^1)$ coefficient, it then follows that
\begin{equation}
\ln {\cal Z}=\sum_D \widetilde{C}(D){\cal F}(D)\,,
\label{lnZ}
\end{equation}
where $D$ is a diagram in the replicated theory with kinematic part 
${\cal F}(D)$, and $\widetilde{C}(D)$ is the ${\cal O}(N^1)$ part of $C_N(D)$, the 
colour factor of $D$ in the replicated theory. 
Eq.~(\ref{lnZ}) gives the sought exponent; in other words we have obtained: 
\begin{equation}
{\cal Z}=\exp\left[\sum_D \widetilde{C}(D){\cal F}(D)\right]\,.
\label{Zexp}
\end{equation}
Equation~(\ref{Zexp}) explicitly expresses the exponentiation of soft gluons, 
and provides a means to calculate the exponent directly: one draws
all possible diagrams in the replicated theory, and calculates the part 
of each diagram which
is linear in the number of replicas $N$. This then enters the exponent
according to eq.~(\ref{Zexp}). 

\begin{figure}
\begin{center}
\scalebox{0.8}{\includegraphics{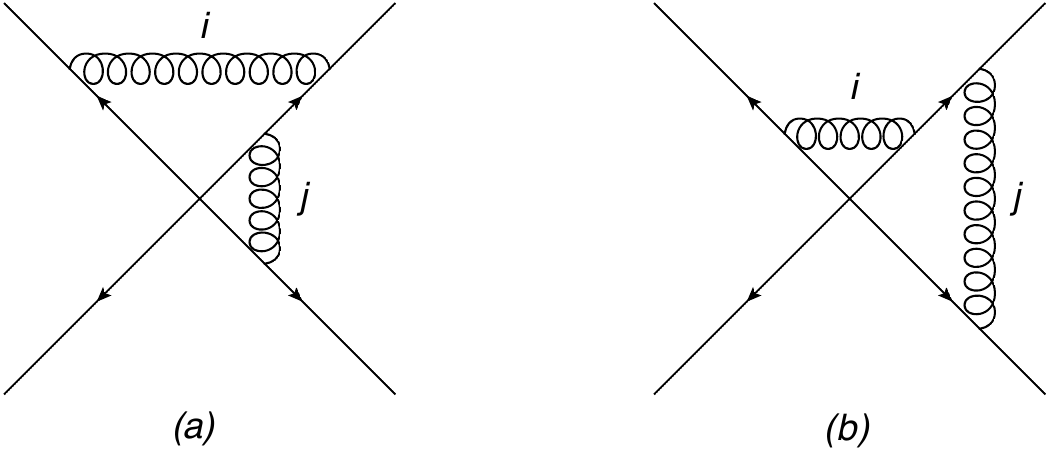}}
\caption{Example diagrams in the replicated theory, where the indices label
the replica number of each gluon.}
\label{twoemfig}
\end{center}
\end{figure}

Many examples of this procedure were considered 
in~\cite{Gardi:2010rn}. To illustrate it, we repeat here the simple two-loop example shown 
in figure~\ref{twoemfig}, which shows two diagrams in the replicated theory,
where each gluon has an associated replica index ($i$ and $j$, respectively). 
The two diagrams of figure~\ref{twoemfig} have the conventional colour factors (the colour factors in the original theory):
\begin{subequations}
\begin{align}
C(a)=T^A\otimes T^BT^A\otimes T^B\otimes I\,;
\label{C(a)}
\\
C(b)=T^A\otimes T^AT^B\otimes T^B\otimes I\,,
\label{C(b)}
\end{align}
\end{subequations}
where $T$ is the colour generator in the representation corresponding to a given eikonal line (which we do not specify here) and $A$ and $B$ denote adjoint indices: as usual, these are summed over for any gluon exchange. 

Consider now diagram $(a)$ in the replicated theory. In order to obtain its colour factor $C_N({\rm a})$ one must consider all possible assignments of replica numbers to the two gluons.
Thus the two replica indices $i$ and $j$ vary, independently, between 1 and $N$.
The total colour factor has contributions from $i=j$, $i<j$ and $i>j$. The first
of these contributes $NC(a)$
i.e. the colour factor is the same as in the original theory, and there is
a multiplicity factor of $N$ corresponding to the number of ways of choosing 
$i=j$. If $i<j$, the colour matrices are reordered on the second parton line, 
corresponding to the fact that the replica numbers must increase away from
the hard interaction; this therefore yields the conventional colour factor of diagram $(b)$. Thus, summing over $i$ and $j$ with $i<j$ one gets a total contribution
\begin{displaymath}
\frac{N(N-1)}{2}C(b),
\end{displaymath}
where the prefactor arises from the number of ways of choosing $i<j$.
Finally, if $i>j$ the replica numbers are already ordered on the second
parton line, and one gets a contribution
\begin{displaymath}
\frac{N(N-1)}{2}C(a).
\end{displaymath}
Putting things together, the total colour factor of diagram $(a)$ in the replicated theory, which we denote by $C_N(a)$, is
\begin{align}
C_N(a)&=NC(a)+\frac{N(N-1)}{2}C(b)+\frac{N(N-1)}{2}C(a)\notag\\
&=\frac{N}{2}[C(a)-C(b)]+\frac{N^2}{2}[C(a)+C(b)].
\label{hatCa}
\end{align}
The replica trick then tells us that this diagram contributes to the exponent
of the Wilson-line correlator with an exponentiated colour factor
\begin{equation}
\widetilde{C}(a)=\frac{1}{2}[C(a)-C(b)],
\label{tildeCa}
\end{equation}
which is the ${\cal O}(N^1)$ part of $C_N(a)$ in eq.~(\ref{hatCa}). Note that this is a
linear combination of the original colour factors of both diagrams in 
figure~\ref{twoemfig}. Carrying out a similar procedure for 
figure~\ref{twoemfig}$(b)$, one finds
\begin{equation}
\widetilde{C}(b)=\frac{1}{2}[C(b)-C(a)],
\label{tildeCb}
\end{equation}
and thus the diagrams of figure~\ref{twoemfig} form a closed set, mixing
only with each other under exponentiation. The
contribution of these two diagrams to the Wilson-line-correlator exponent is
\begin{align}
\widetilde{C}(a){\cal F}(a)+\widetilde{C}(b){\cal F}(b)=
\left(\begin{array}{c}{\cal F}(a)\\{\cal F}(b)\end{array}\right)^T
\left(\begin{array}{c}\widetilde{C}(a)\\\widetilde{C}(b)\end{array}\right)
=\left(\begin{array}{c}{\cal F}(a)\\{\cal F}(b)\end{array}\right)^T
\frac{1}{2}\left(\begin{array}{rr}1&-1\\-1&1\end{array}\right)
\left(\begin{array}{c}C(a)\\C(b)\end{array}\right),
\end{align}
which agrees with the explicit calculation presented in Ref.~\cite{Aybat:2006mz}.
Comparing this to eq.~(\ref{exp2}), one sees that the web mixing matrix
associated with this pair of diagrams is
\begin{equation} 
R=\frac{1}{2}\left(\begin{array}{rr}1&-1\\-1&1\end{array}\right)\,.
\label{Rtwoem}
\end{equation}
It is easy to verify that $R$ indeed has the two properties of eqs.~(\ref{idemp}) 
and~(\ref{zerosum}). 

This simple example illustrates why diagrams at any loop order form closed 
sets. For our purposes in the next section it is useful to note that one may systematise the replica ordering operation 
above by momentarily assigning replica indices to the colour generators associated with
each gluon emission, then ordering the generators as required by the ordering operator ${\cal R}$, and finally removing these indices.  
For example, the colour factor of eq.~(\ref{C(a)}) may be written in the replicated theory as
\begin{equation}
C_N(a)=\sum_{i,j=1}^N{\cal R}\left[T_i^A\otimes T_j^BT_i^A\otimes 
T_j^B\otimes I\right],
\label{hatCa2}
\end{equation}
where $T_i^A$ is a colour generator with adjoint index $A$ and replica
number $i$. The sum is over all assignments of replica numbers, and one may
separate this sum into the three hierarchies of replica numbers given above
i.e.
\begin{equation}
C_N(a)=\left[\sum_{i=j}+\sum_{i<j}+\sum_{i>j}\right]{\cal R}
\left[T_i^A\otimes T_j^BT_i^A\otimes T_j^B\otimes I\right].
\label{hatCa3}
\end{equation}
One may now use the fact that
\begin{equation}
{\cal R}\left[T_i^A\otimes T_j^BT_i^A\otimes T_j^B\otimes I\right]=
\left\{\begin{array}{l}T_i^A\otimes T_i^AT_j^B\otimes T_j^B\otimes I,\quad i<j\\
T_i^A\otimes T_j^BT_i^A\otimes T_j^B\otimes I,\quad{\rm otherwise}
\end{array}\right.
\end{equation}
to rewrite eq.~(\ref{hatCa3}) as
\begin{equation}
C_N(a)=\left[\sum_{i=j}+\sum_{i>j}\right]
\left[T_i^A\otimes T_j^BT_i^A\otimes T_j^B\otimes I\right]+\sum_{i<j}
\left[T_i^A\otimes T_i^AT_j^B\otimes T_j^B\otimes I\right].
\label{hatCa4}
\end{equation}
Having carried out replica ordering, the replica indices may now be removed.
Each distinct sum then gives a multiplicity factor times the appropriate
colour factor, so that eq.~(\ref{hatCa4}) becomes
\begin{equation}
C_N(a)=\left[N+\frac{N(N-1)}{2}\right]C(a)+\frac{N(N-1)}{2}C(b),
\label{hatCa5}
\end{equation}
in agreement with eq.~(\ref{hatCa}) above. In general, the colour factor of a 
diagram $D$ in the replicated theory has a form similar to eq.~(\ref{hatCa2}), with a sum over
all possible replica numbers and a string of colour matrices associated with
each Wilson line. 
The ${\cal R}$ operation then interchanges colour matrices on
each line, such that the colour factor in the replicated theory is given by
a superposition of the conventional colour factors of all graphs related to 
the original graph by gluon permutations along each of the Wilson lines. 
That is, eq.~(\ref{Ctilde}) holds, where $D'$ runs over the closed set of diagrams obtained by taking diagram $D$ and permuting the gluons on each external line. The above procedure of
assigning replica numbers to colour generators will be useful when proving 
the idempotence property of the web mixing matrices in section~\ref{sec:idemp}.

In this section, we have reviewed the replica trick formalism
of~\cite{Gardi:2010rn} for calculating the exponent of the soft gluon
amplitude. In the following section we will see that a similar line of argument
can be used to prove the idempotence of web mixing matrices. 

%%%%%%%%%%%%%%%%%%%%%%%%%%%%%%%%%%%%%%%%%%%
\section{Proof of idempotence}
\label{sec:idemp}

In section~\ref{sec:replica} we reviewed the use of the replica trick 
in deriving the exponent of a correlator of any number of Wilson lines, recalling how
web mixing matrices arise and how they are computed. In this section we will extend this
argument to prove the idempotence of mixing matrices.

Our starting point is the generating functional for the original theory,
given by eq.~(\ref{Zdef2}). As already shown in eq.~(\ref{Zrepdef2}), we
may replicate the theory to produce a generating functional 
\begin{align}
{\cal Z}^M&=\int[{\cal D}A_1^\mu]\ldots[{\cal D}A_M^\mu]
\, \,{\rm e}^{{\mathrm i}\sum_{i=1}^M S[A_i^\mu]}\,\,\times
\notag\\&\hspace*{50pt}
{\cal R}\left\{{\cal P}
 \exp\left[\sum_{i=1}^M\int dt
\beta_1\cdot A_i\right]\otimes\,\ldots\,\otimes {\cal P}\exp\left[\sum_{i=1}^M\int dt
\beta_L\cdot A_i\right]\right\},
\label{ZrepdefM}
\end{align}
where the ${\cal R}$ operator orders gluon emissions by replica number,
such that the latter increases away from the hard interaction.
Here we have used $M$ as the number of replicas, for reasons
that will become clear. We may shorten notation in eq.~(\ref{Zrepdef2}) 
by introducing the following definitions:
\begin{subequations}
\begin{align}
{\cal D}A_\mu^{(I)}&={\cal D}{A_1}_\mu{\cal D}{A_2}_\mu\ldots{\cal D}{A_{M}}_\mu;
\label{DAprod}\\
S[A_\mu^{(I)}]&=\sum_{i=1}^MS[{A_i}_\mu];\label{Ssum}\\
\Phi_{k}^{(I)}&={\cal P}\exp\left({\mathrm i}\sum_{i=1}^M
\int dx_k^\mu {A_i}_\mu\right).\label{Wilprod}
\end{align}
\end{subequations}
That is, we associate a superindex $I$ with the entire set of replica 
numbers from $1$ to $M$, such that eq.~(\ref{Zrepdef2}) becomes
\begin{equation}
{\cal Z}^M=\int{\cal D}A_\mu^{(I)}\,{\rm e}^{\,{\mathrm i}S[A_\mu^{(I)}]}
{\cal R}\left[\Phi_1^{(I)}\otimes\ldots\otimes\Phi_L^{(I)}\right].
\label{Zrepdef3}
\end{equation}
This is highly suggestive, as it looks schematically just like the 
original theory defined by eq.~(\ref{Zdef}). The only difference is
the presence of the ${\cal R}$ operator, which modifies the colour 
factors as we have already discussed, such that the ${\cal O}(M^1)$
part of the colour factor of any given diagram is its
ECF rather than the conventional colour factor. 
Given that ${\cal R}$ acts separately
on each parton line, we may rewrite eq.~(\ref{Zrepdef3}) slightly to give
\begin{equation}
{\cal Z}^M=\int{\cal D}A_\mu^{(I)}\,{\rm e}^{{\mathrm i}S[A_\mu^{(I)}]}
\left[\left({\cal R}\Phi_1^{(I)}\right)\otimes\,\ldots\,\otimes \left({\cal R}\Phi_M^{(I)}\right)
\right].
\label{Zrepdef4}
\end{equation}
>From now on, we refer to this as the generating functional for the 
{\it singly replicated} theory. That is, we have applied the replica
trick once to the original (non-replicated) theory. This yields singly-replicated colour factors, $C_M(D)$ whose ${\cal O}(M^1)$ parts are the ECF's $\widetilde{C}(D)$; these determine the exponent in the original theory:
\begin{equation}
\ln {\cal Z}=\sum_D \widetilde{C}(D){\cal F}(D)\,.
\label{lnZ_singly}
\end{equation}

Having constructed the singly-replicated theory described by 
eq.~(\ref{Zrepdef4}), our next step is to construct the generating
functional
\begin{align}
\label{ZrepN}
\begin{split}
\left(Z^{M}\right)^N&=\int{\cal D}A_\mu^{(I_1)}\ldots{\cal D}A_\mu^{(I_N)}
\,\,{\rm e}^{{\mathrm i}\sum_{j=1}^N S[A_\mu^{(I_j)}]}\,\,\times
\\&\hspace*{10pt}
\left[
\left({\cal R}\Phi_1^{(I_1)}\right)
\left({\cal R}\Phi_1^{(I_2)}\right)
\ldots
\left({\cal R}\Phi_1^{(I_N)}\right)\otimes\,\ldots
\,\otimes
\left({\cal R}\Phi_L^{(I_1)}\right)
\left({\cal R}\Phi_L^{(I_2)}\right)
\ldots\left({\cal R}\Phi_L^{(I_N)}\right)
\right].
\end{split}
\end{align}
Here, we have replicated the singly-replicated theory by creating $N$ 
identical copies of each block of $M$ replica numbers. We may take
the replica numbers of individual gluons in block $j$ as going from
$[(j-1)M+1]$ to $jM$, and there are $NM$ individual replicas of the original theory in
total. Associated with each block of replicas $j$ is
\begin{itemize}
\item A path integral over the set of gauge fields $A_\mu^{I_j}$.
\item A term in the action $S[A_\mu^{I_j}]$; as usual these combine additively
due to the fact that the replicas do not interact with each other.
\item A replica-ordered factor $\left({\cal R}\Phi_k^{(I_j)}\right)$ on parton 
line-$k$, where the replica-ordering operator is associated with the first replication, thus
acting only within the specific block of replica numbers associated with $I_j$.
\end{itemize}
Clearly the generating functional of eq.~(\ref{ZrepN}) is related to 
that of eq.~(\ref{Zrepdef4}) by being the latter raised to the 
$N^{\rm th}$ power, as recognized on the left-hand side of 
eq.~(\ref{ZrepN}). We will call this the generating functional for the 
{\it doubly-replicated theory}, due to the fact that the replica trick
has now been applied twice - once in replicating gluons in the original
theory (to make the singly-replicated theory), and again in replicating
the blocks of replicas (to make the doubly-replicated theory).
In any diagram generated by (\ref{ZrepN}) the \emph{block indices} $I_j$ must be increasing as one moves away from the hard interaction, analogously to the way that replica 
indices must increase in the singly-replicated theory. 

As in the singly-replicated theory, one may write the Wilson line factors
in eq.~(\ref{ZrepN}) as a single Wilson line factor. To do this one inserts
an additional replica-ordering operator~${\cal R}$, so as to 
write the Wilson-line factors on line $k$ as
\begin{equation}
\left({\cal R}\Phi_k^{(I_1)}\right)\left({\cal R}\Phi_k^{(I_2)}\right)\ldots\left({\cal R}\Phi_k^{(I_N)}\right)
={\cal R}{\cal R}{\cal P}\exp\left[i\sum_{j=1}^N\int
dx_k^\mu A_\mu^{(I_j)}\right].
\label{Wilrep}
\end{equation}
Here the replica-ordering operator that acts first (closest to the Wilson line)
orders the replica numbers within each block. The second operator acts
to order the block indices along the Wilson line. The generating 
functional for the doubly-replicated theory, equation~(\ref{ZrepN}),
can then be written as
\begin{align}
\label{ZrepN2}
\begin{split}
{\cal Z}^{MN}&=\int{\cal D}A_\mu^{I_1}\ldots{\cal D}A_\mu^{I_N}\,\,
{\rm e}^{{\mathrm i}\sum_{j=1}^N
S[A_\mu^{(I_j)}]}\,\,\times\\
&\hspace*{50pt}
{\cal R}\left[{\cal R}{\cal P}\exp\left(
i\sum_{j=1}^Ndx_1^\mu A_\mu^{(I_j)}\right)\otimes\,\ldots
\,\otimes {\cal R}{\cal P}\exp\left(
i\sum_{j=1}^Ndx_L^\mu A_\mu^{(I_j)}\right)\right].
\end{split}
\end{align}
Again, the replica-ordering operator inside the square brackets acts within
each block of replica numbers. The replica-ordering operator outside the
square bracket acts to order the block indices. We could have used a different
symbol for each of these operators; however, they result from the same
replica ordering operation as applied to individual gluons, hence the 
use of the same notation.

Let us now examine colour factors in the doubly-replicated theory. 
Using a similar notation to that of section~\ref{sec:replica}, we
denote by $C_{MN}(D)$ the colour factor of diagram $D$ in the 
doubly-replicated theory, and $\widetilde{C}'(D)$ the ${\cal O}((MN)^1)$ part of this. 
A summary of properties of the original, singly-replicated and 
doubly-replicated theories is shown in table~\ref{reptab}.
\begin{table}
\begin{center}
\begin{footnotesize}
\begin{tabular}{|c||c|c|c|}
\hline
Property & Original Theory & Singly-Replicated Theory 
& Doubly-Replicated Theory\\
\hline
Generating functional & ${\cal Z}$ & ${\cal Z}^M$ & ${\cal Z}^{MN}$\\
Result of replication & Singly-replicated theory & Doubly-replicated theory
& ---\\
Colour factors & $C(D)$ & $C_M(D)$ & $C_{MN}(D)$\\
ECF's & $\widetilde{C}(D)=\left.C_M(D)\right\vert_{{\cal O}(M^1)}$ & $\widetilde{C}'(D)=
\left.C_{MN}(D)\right\vert_{{\cal O}((MN)^1)}=\widetilde{C}(D)$& --- \\
Action of ${\cal R}$ & ---&Ordering of replica indices & 
Ordering of block indices\\
\hline
\end{tabular}
\end{footnotesize}
\caption{Table representing the relationships between the original and
replicated theories. The fact that the ECF's of the singly-replicated theory $\widetilde{C}'(D)$ coincide with the ECF's of the original theory $\widetilde{C}(D)$ will be derived below, see eq.~(\ref{C_eq_C'}).}
\label{reptab}
\end{center}
\end{table}

By the usual replica trick argument, one has
\begin{equation}
{\cal Z}^{MN}=1+NM\log({\cal Z})+{\cal O}(M^2N^2),
\label{doublerep}
\end{equation}
and thus that the generating functional of the original (non-replicated)
theory exponentiates, taking the form:
\begin{equation}
\ln {\cal Z}=\sum_D \widetilde{C}'(D){\cal F}(D)\,,
\label{lnZ_doubly}
\end{equation}
where the colour factors in the exponent are given by the ${\cal O}((MN)^1)$ parts of the colour factors in the doubly-replicated theory, $\widetilde{C}'(D)$. 
We can find these by using a similar procedure to that presented
in section~\ref{sec:replica}. First, one considers the string of colour matrices 
associated with each external line, and assigns two sets of indices to each 
generator. For example, considering a line with $s$ gluon emissions, we may 
write the associated string of colour generators as
\begin{equation}
T^{I_1}_{i_1}\ldots T^{I_s}_{i_s},
\label{colstring}
\end{equation}
where we have suppressed the adjoint indices. The subscripts in eq.~(\ref{colstring})
denote replica indices taking values from $1\ldots M$, whereas the superscripts are
block indices ranging from $1\ldots N$, which indicate which block of $M$ replicas a given 
gluon originates from. In the doubly-replicated theory, the procedure for finding the colour
factor of a graph is as follows:
\begin{enumerate}
\item Write down the conventional colour factor of the graph (as 
in the non-replicated theory).
\item Consider all possible assignments of replica indices. For each such 
assignment, reorder the colour matrices according to the replica-ordering 
operator. 
\item At this stage, the replica indices can be removed, and 
the ${\cal O}(M)$ part of the colour factor of the graph is
\begin{equation}
\widetilde{C}(D)=\sum_{D'} R_{DD'}C(D'),
\label{widetildeCD}
\end{equation}
as in section~\ref{sec:replica}.
\item Next, consider all possible assignments of block indices. For each assignment, 
reorder the colour matrices according to the block index ordering operator, the second ${\cal R}$ operator in eq.~(\ref{ZrepN2}). 
The ordering of block indices is exactly analogous to the ordering of replica indices, and introduces \emph{the same} mixing matrix $R$ again, only now it acts on $\widetilde{C}$, since the internal indices within each block have already been ordered in the previous steps.
\item 
Now that the blocks have been ordered, the block indices can also be removed, and the ${\cal O}(N)$ part of the colour factor is
\begin{equation}
\widetilde{C}'(D)=\sum_{E}R_{DE}\widetilde{C}(E)=\sum_{E,D'}R_{DE}R_{ED'}C(D')\,,
\label{Ctilde'}
\end{equation} 
where in the last step we used (\ref{widetildeCD}).
\end{enumerate}
Note that in the above procedure that we ordered the replica indices (associated with $M$) followed
by the block indices (associated with $N$). However, this order is unimportant, due to the fact
that the ${\cal O}(MN)$ part of the colour factor in the 
doubly-replicated theory can be obtained
by taking the ${\cal O}(N)$ part of the ${\cal O}(M)$ part (as done here), or vice versa. 

At first sight one may consider $\widetilde{C}'(D)$ in (\ref{Ctilde'}) to be different from $\widetilde{C}(D)$. In fact they are the same. This follows straightforwardly from eq.~(\ref{doublerep}): as summarized by (\ref{lnZ_doubly}), the original generating functional ${\cal Z}$ exponentiates, with colour factors which are the ${\cal O}((MN)^1)$ parts of the colour factors in the doubly-replicated theory ($\widetilde{C}'(D)$). 
However, we already know from the original replica trick argument, that ${\cal Z}$ exponentiates with ECF's $\widetilde{C}(D)$, eq.~(\ref{lnZ_singly}) above. 
Thus we deduce that 
\begin{equation}
\label{C_eq_C'}
\widetilde{C}(D)=\widetilde{C}'(D)\,, 
\end{equation}
as anticipated in table~\ref{reptab}.

The essential reason for this can be seen as follows. In eq.~(\ref{ZrepN2})
we introduced two replica ordering operators, where the first orders replica 
indices within each block of $M$ replica numbers, and the second orders the $N$
blocks in sequence. There is, however, a second way to think about the doubly
replicated theory, namely as a singly replicated theory with $NM$ individual
replica numbers, with the replica indices in block $j$ going from
$[(j-1)M+1]$ to $jM$ as discussed above. The double replica ordering of 
block indices and replica indices within blocks is then entirely equivalent
to replica ordering of the $NM$ individual replica numbers. That is, one
may write
\begin{equation}
{\cal R}^2={\cal R}
\label{Ridemp}
\end{equation}
in eq.~(\ref{ZrepN2}), where ${\cal R}$ on the right-hand side is the usual
replica ordering operator. The generating functional for the doubly replicated
theory can then be rewritten
\begin{align}
\label{ZrepN3}
\begin{split}
{\cal Z}^{NM}&=\int{\cal D}A_\mu^{I_1}\ldots{\cal D}A_\mu^{I_N}\,\,
{\rm e}^{{\mathrm i}\sum_{j=1}^N S[A_\mu^{(I_j)}]}\,\,\times\\
&\hspace*{30pt}{\cal R}\left[{\cal P}\exp\left(
{\mathrm i}\sum_{j=1}^Ndx_1^\mu A_\mu^{(I_j)}\right)\otimes\,\ldots\,\otimes {\cal P}\exp\left(
{\mathrm i}\sum_{j=1}^Ndx_L^\mu A_\mu^{(I_j)}\right)\right].
\end{split}
\end{align}
Now we may replace each superindex $I_j$ with its corresponding block of
$M$ individual replica numbers, so that eq.~(\ref{ZrepN3}) becomes
\begin{align}
\label{ZrepN4}
\begin{split}
{\cal Z}^{NM}&=\int{\cal D}A_\mu^{(1)}\ldots{\cal D}A_\mu^{(MN)}\,\,
{\rm e}^{{\mathrm i}\sum_jS[A_{j\mu}]}\,\,\times\\
&\hspace*{30pt}
{\cal R}\left[{\cal P}\exp\left(
{\mathrm i}\sum_{j=1}^{MN}dx_1^\mu A_{j\mu}\right)\otimes\,\ldots\,\quad\otimes {\cal P}\exp\left(
{\mathrm i}\sum_{j=1}^{MN}dx_L^\mu A_{j\mu}\right)\right]\,.
\end{split}
\end{align}
This has manifestly the same form as the singly-replicated theory given by 
eq.~(\ref{Zrepdef}), but where there are $MN$ replica numbers rather than
$M$. That is, upon replicating the singly-replicated theory, one obtains 
a theory that is equivalent to a singly-replicated theory. In other words,
one does not gain anything by replicating the theory twice, due
to the idempotence of the replication procedure. This is ultimately a consequence of the fact that the replica operation ${\cal R}$ is itself idempotent. 

We have shown that the colour factors $\widetilde{C}'(D)$ and $\widetilde{C}(D)$ are equal.
Combining this information with eqs.~(\ref{Ctilde}) and~(\ref{Ctilde'}) gives
\begin{equation}
\sum_{E,D'}R_{DE}R_{ED'}C(D')=\sum_{D'}R_{DD'}C(D')
\label{idemp1}
\end{equation}
and thus
\begin{equation}
R_{DD'}=\sum_{E,D'}R_{DE}R_{ED'},
\label{idemp2}
\end{equation}
which is the idempotence property of eq.~(\ref{idemp}). This completes the proof. 

The interpretation of the idempotence property has already been discussed in~\cite{Gardi:2010rn}: the mixing matrix $R$ is a projection operator.
An important corollary is that $R$ is diagonalizable, with eigenvalues which are either 0 or 1. 
By rewriting eq.~(\ref{exp2}) in the diagonal basis, Ref.~\cite{Gardi:2010rn} showed that the exponent is built exclusively out of the linear combinations of colour and kinematic factors corresponding to the eigenvalue $1$, while the entire subspace\footnote{It should be noted that this subspace is never empty, as guaranteed by the zero sum row property~(\ref{zerosum}).} corresponding to eigenvalue $0$ is removed. This structure was then linked to the cancellation of subdivergences which must occur for webs to conform with the  
renormalization of the multieikonal vertex~\cite{Mitov:2010rp,Gardi:2010rn}.
The proof of idempotence is thus an important step in understanding the structure of infrared singularities.

In this section, we have proved the idempotence property of web mixing matrices. 
The proof proceeds by applying the replica trick twice to the original soft gluon
theory. Idempotence of the mixing matrices then stems from the fact that a doubly-replicated theory is equivalent to a singly-replicated one.
The above analysis demonstrates the power of the replica trick formalism: not only does it facilitate a direct calculation of the exponent, it also provides deep insight into its structure.

%%%%%%%%%%%%%%%%%%%%%%%%%%%%%%%%%%%%%%%%%%%%%%%%
\section{Combinatoric formulae for webs in terms of overlap functions}
\label{sec:comb}

In the previous sections, we reviewed the replica trick formalism which determines the structure of webs and then used it to prove the idempotence property of the mixing matrices.
One can in fact go further and find explicit combinatoric formulae relating the exponentiated
colour factors $\widetilde{C}(D)$ to the conventional colour factors $C(D)$. Such formulae were first presented in section 4.1 in \cite{Gardi:2010rn}; here we develop these further so as to establish an explicit expression for the mixing matrix elements in terms of combinatoric ``overlap functions''. This will give us a handle to study further properties of these matrices, and in particular, to prove the zero sum row property of eq.~(\ref{zerosum}), which we address in the next section.   

\begin{figure}
\begin{center}
\scalebox{0.8}{\includegraphics{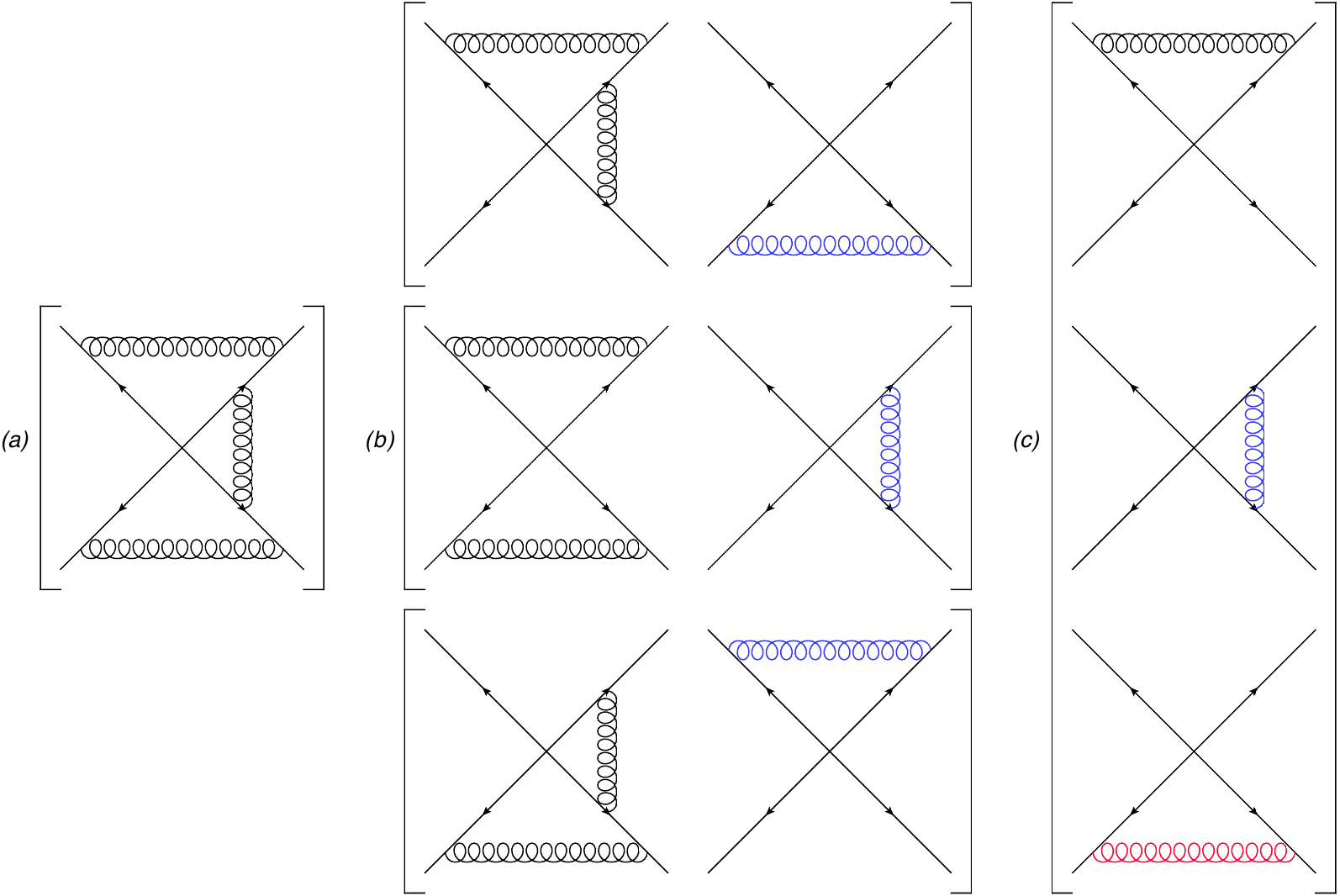}}
\caption{Decompositions of an example three loop diagram, with (a) $n(P)=1$
(equivalent to the original diagram); (b) $n(P)=2$; (c) $n(P)=3$.}
\label{partex2}
\end{center}
\end{figure}

First, we introduce the notion of a {\it decomposition} $P$ of a diagram $D$.
This is a partitioning of $D$ into a number of parts, each
containing one or more connected pieces from $D$. One may label a given
decomposition by the number of elements $n(P)$ it contains\footnote{Note that there may be more than one decomposition with a given number of elements. For example in figure~\ref{partex2} there are three distinct decompositions $P$ having $n(P)=2$.}. An example of a graph
together with its decompositions is shown in figure~\ref{partex2}.
In the replica trick formalism, decompositions arise as possible assignments of
replica numbers to a given graph. An assignment with $m$ distinct replica 
numbers corresponds to a decomposition with $n(P)=m$, where each element
in the decomposition corresponds to a single replica number. We have 
represented different replica numbers by different colours in 
figure~\ref{partex2}. Each decomposition consists of a set of subdiagrams
$g_1,g_2\ldots,g_{n(P)}$, and in~\cite{Gardi:2010rn} it was shown that
the ECF of any diagram $D$ can be written in terms of the conventional
colour factors of its possible subdiagrams as follows:
\begin{equation}
\widetilde{C}(D)=\sum_P\frac{(-1)^{n(P)-1}}{n(P)}\sum_\pi C(g_{\pi_1})
\ldots C(g_{\pi_{n(P)}}),
\label{modcol2}
\end{equation}
where the sum is over all permutations $\pi$ of $[1,\,2,\,\ldots\, n(P)]$. In each
term, the ordering of the colour factors is important owing to the fact 
that they do not commute. 

\begin{figure}
\begin{center}
\scalebox{0.8}{\includegraphics{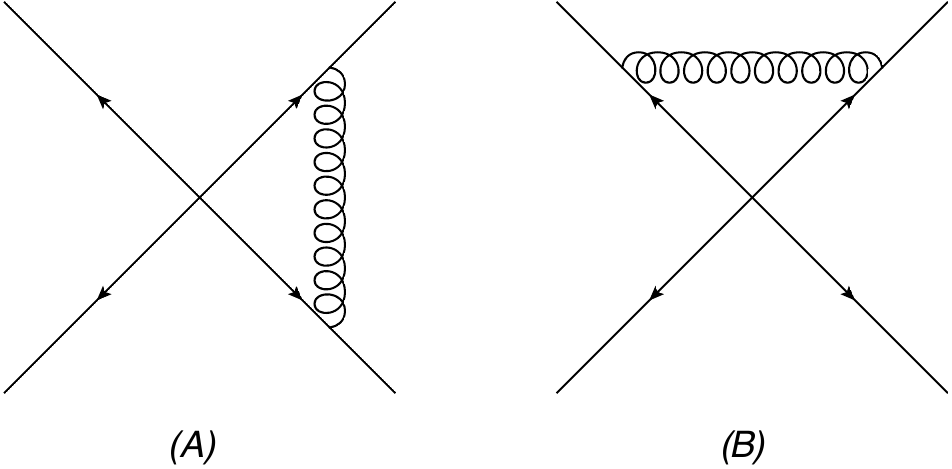}}
\caption{Example one loop graphs.}
\label{1ggraphs}
\end{center}
\end{figure}

\begin{figure}[htb]
\begin{center}
\scalebox{0.8}{\includegraphics{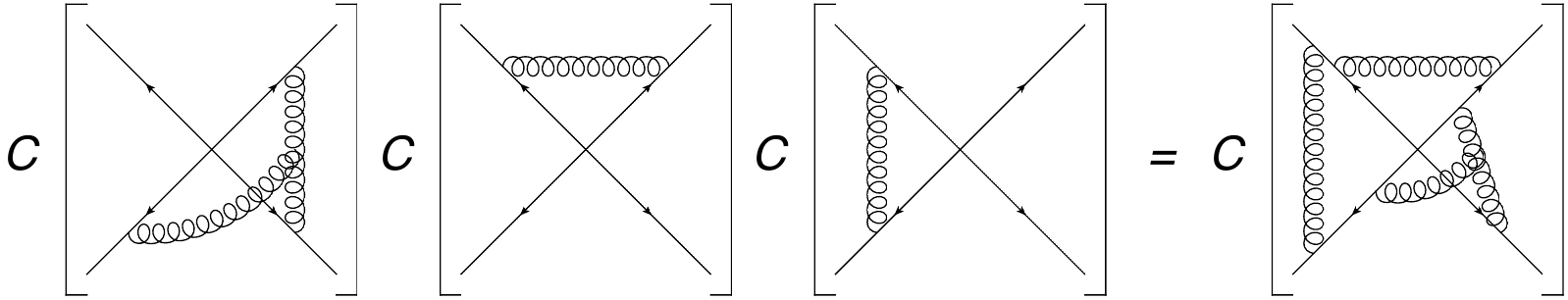}}
\caption{Example of a product of colour factors of three diagrams.}
\label{prodex}
\end{center}
\end{figure}

Our next step will be to extract an explicit 
combinatoric expression for the mixing matrix element $R_{DD'}$ from 
eq.~(\ref{modcol2}). Each term on the right-hand side of this equation contains
a product of conventional colour factors of subdiagrams of the diagram $D$ on the 
left-hand side. The rule for dealing with products of colour factors is as 
follows. If the colour factors of two diagrams $G$ and $H$ involve strings of
generators on a given parton line
\begin{displaymath}
T^{A_1}\ldots T^{A_{s_1}}
\end{displaymath}
and
\begin{displaymath}
T^{B_1}\ldots T^{B_{s_2}}
\end{displaymath}
respectively, where $A_{s_i}$ and $B_{s_i}$ are adjoint indices, 
then the colour factor $C(G)C(H)$ involves the string
\begin{displaymath}
T^{A_1}\ldots T^{A_{s_1}}T^{B_1}\ldots T^{B_{s_2}}.
\end{displaymath}
One then applies this rule to every parton line in turn. As an example,
consider the two graphs shown in figure~\ref{1ggraphs}. 
These diagrams have colour factors
\begin{subequations}
\begin{align}
C(A)&=I\otimes T^A\otimes T^A \otimes I
\label{C(A)}
\\
C(B)&=T^B\otimes T^B\otimes I\otimes I,
\label{C(B)}
\end{align}
\end{subequations}
where, as usual, $I$ is the identity matrix in colour space. 
The product of colour factors is
\begin{equation}
C(A)C(B)=T^B\otimes T^AT^B\otimes T^A\otimes I,
\end{equation}
which we recognise (from eq.~(\ref{C(a)}) after relabelling) as the 
colour factor of the diagram in figure~\ref{twoemfig}(a). 
As perhaps is clear from this example, there is a simple graphical 
interpretation for the product of colour factors. The colour factor of 
a product of subdiagrams is the colour factor of the graph obtained
by drawing each subdiagram in sequence, moving progressively outwards
from the hard interaction vertex. A more complicated example is shown
in figure~\ref{prodex}.

Returning to eq.~(\ref{modcol2}), each term on the right-hand side contains
\emph{a complete set} of subdiagrams of $D$, by which we mean that all connected
pieces of $D$ occur in one of the subdiagrams on the right-hand side. 
The product of the colour factors of the subdiagrams of $D$ in a given decomposition 
gives either the colour factor of $D$ itself, or one of the other diagrams in the closed set -- web -- to which $D$ belongs (it can only differ from $D$ by permutation of the gluon attachments to the Wilson lines).
Furthermore, each product uniquely specifies a diagram in the set. 
That is,
\begin{equation}
C(g_{\pi_1})\ldots C(g_{\pi_{n(P)}})=C(D')
\label{prodD'}
\end{equation}
for some $D'$ related to $D$ by permutations of gluons on the external lines. 
Let $P_D$ denote a given decomposition of $D$. Then we may write
\begin{equation}
\sum_\pi C(g_{\pi_1})\ldots C(g_{\pi_{n(P_D)}})=\sum_{D'}\ip{D'}{P_D}C(D'),
\label{sumpi}
\end{equation}
where we have introduced $\ip{D'}{P_D}$, which we call the 
{\it overlap of $D'$ with the decomposition $P_D$ of $D$}. 
That is, the number of ways that diagram $D'$ is formed by considering all the permutations
of all the elements in $P_D$. 

\begin{figure}[htb]
\begin{center}
\scalebox{.93}{\includegraphics{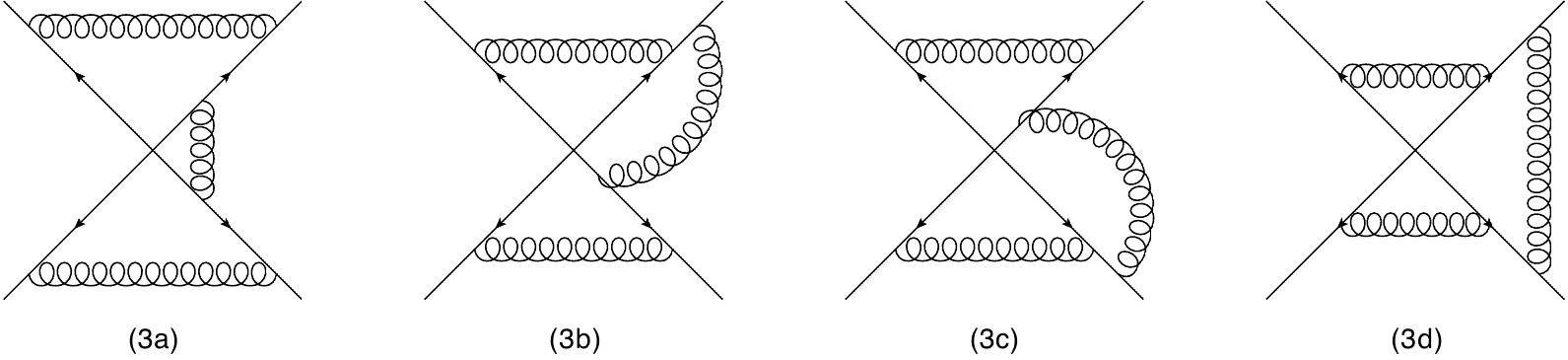}}
\caption{Three loop web to which the diagram of figure~\ref{partex2}(a) belongs.}
\label{17-20}
\end{center}
\end{figure}

To clarify the notation let us examine a couple of examples.
Consider the set of graphs given in figure~\ref{17-20}, 
which constitute the web to which the diagram of figure~\ref{partex2}(a) belongs, 
where we have labelled the diagrams as in~\cite{Gardi:2010rn}. 
Now consider the decomposition of diagram (3a) given by the top diagram in 
figure~\ref{partex2}(b). There are two permutations of the elements of this decomposition.
Using the graphical rule exemplified by figure~\ref{prodex}, one sees that one
permutation makes diagram (3a), and the other makes diagram (3c). That is, labelling
this decomposition by $P$, one has
\begin{equation}
\ip{(3a)}{P}=1,\quad\ip{(3b)}{P}=0,\quad\ip{(3c)}{P}=1,\quad \ip{(3d)}{P}=0.
\label{ips}
\end{equation}

As a second example, consider the decomposition of (3a) shown in 
figure~\ref{partex2}(c), consisting of three distinct replica numbers,
and which we label by $Q$.
There are six permutations of the elements of the decomposition, and
the reader may verify that
\begin{equation}
\ip{(3a)}{Q}=2,\quad\ip{(3b)}{Q}=1,\quad\ip{(3c)}{Q}=1,\quad\ip{(3d)}{Q}=2.
\label{ips2}
\end{equation}

As argued above, each permutation of the elements of a decomposition
$P_D$ of a given diagram $D$ contributes to one (and only one) diagram $D'$ in 
the web to which $D$ belongs. Different permutations may, of course, 
give the same diagram $D'$, as occurs in the example of eq.~(\ref{ips2})
above, in which diagrams (3a) and (3d) are each formed from two different
permutations. It follows that if one sums the overlap functions for a 
given decomposition over all diagrams in the web, this must be equal to the
total number of permutations in the decomposition $P_D$. That is,
\begin{equation}
\sum_{D'}\ip{D'}{P_D}=n(P_D)!,
\label{D'fact}
\end{equation}
where $n(P_D)$ is the number of elements in $P_D$. We will use this 
property in what follows, but note for now that this is indeed satisfied
in eqs.~(\ref{ips}) and~(\ref{ips2}).

Having defined the overlap functions, we may substitute eq.~(\ref{sumpi})
into eq.~(\ref{modcol2}) to obtain
\begin{align}
\widetilde{C}(D)&=\sum_{P_D}\frac{(-1)^{n(P_D)-1}}{n(P_D)}
\sum_{D'}\ip{D'}{P_D}C(D')\notag\\
&=\sum_{D'}\sum_{P_D}\frac{(-1)^{n(P_D)-1}}{n(P_D)}
\ip{D'}{P_D}C(D'),
\label{modcol3b}
\end{align}
where we have interchanged the order of the summations over $D'$ and $P_D$
in the second line. Comparing this with eq.~(\ref{Ctilde}), one finds
\begin{equation}
R_{DD'}=\sum_{P_D}\frac{(-1)^{n(P_D)-1}}{n(P_D)}\ip{D'}{P_D}.
\label{RDD}
\end{equation}
This is an explicit combinatoric formula for the web mixing matrix, which 
should prove useful in further studies of webs. Furthermore, 
eq.~(\ref{RDD}) makes explicit the fact that there are two\footnote{A third layer of combinatoric
complexity occurs when webs are renormalized, involving nested commutators
of lower-order counterterms and webs as explained in~\cite{Mitov:2010rp}.} sources of
combinatoric complexity involved in the structure of the exponent of Wilson-line correlators. 
Firstly, there are combinatoric factors resulting
from exponentiation --- the factor involving $n(P_D)$ in
eq.~(\ref{RDD}). Secondly, there are combinatoric factors relating
to how diagrams in a web are related to one another via permutations of
gluons --- the overlap functions.

%%%%%%%%%%%%%%%%%%%%%%%%%%%%%%%%%%%%%%%%%%%
\section{Proof of the zero sum row property}
\label{sec:zerosum}

Armed with eq.~(\ref{RDD}), which is an explicit formula for the web mixing matrix $R$,
we can now address the zero sum row property. Summing the elements in row $D$ of $R$ 
using (\ref{RDD}) one finds
\begin{equation}
\sum_{D'}R_{DD'}=\sum_{P_D}\frac{(-1)^{n(P_D)-1}}{n(P_D)}\sum_{D'}\ip{D'}{P_D},
\label{sumzero}
\end{equation}
where we have interchanged the order of the summations over $P_D$ and $D'$. We will shortly see that this is indeed zero. First though one may note that, 
given $\ip{D'}{P_D}\geq0$, the right-hand side is able to give zero only through 
cancellations from the alternating signs in the $n(P_D)$-dependent factor. Thus,
there is an interesting interplay between the combinatorics stemming from exponentiation 
(which has been determined using the replica trick), and that coming from the internal structure of the web and encoded in the overlap functions. 

We may simplify eq.~(\ref{sumzero}) using the result of eq.~(\ref{D'fact}), to get
\begin{equation}
\sum_{D'}R_{DD'}=\sum_{P_D}(-1)^{n(P_D)-1}(n(P_D)-1)!
\label{sumzero2}
\end{equation}
Each term in the sum now depends only on the number of elements of each
decomposition $n(P_D)$. We may thus replace the sum over decompositions $P_D$ with a sum
over the number of elements $m=n(P_D)$ in each decomposition, to give
\begin{equation}
\sum_{D'}R_{DD'}=\sum_{m=1}^{n_c}(-1)^{m-1}\,(m-1)!\,N(n_c,m),
\label{sumzero3}
\end{equation}
where $n_c$ is the number of connected pieces in graph $D$, and $N(n_c,m)$ is the 
number of decompositions (of a graph with $n_c$ connected pieces) which have 
$m$ elements. For example, in figure~\ref{partex2} there are three connected
pieces of the full diagram, and thus $n_c=3$. The number of elements in
each decomposition has the range $1<m<3$, and one has $N(3,1)=1$, $N(3,2)=3$
and $N(3,3)=1$, as can be easily verified by counting the number of decompositions
in figures~\ref{partex2}(a), (b) and (c) respectively. 

The problem of proving the zero sum row property now amounts to showing
that the right-hand side of eq.~(\ref{sumzero3}) is zero. To do this, 
note that a decomposition of a given graph is a partition of the set
of its connected pieces into non-empty subsets. Thus, $N(n_c,m)$ counts
the number of partitions of a set of $n_c$ objects into $m$ non-empty 
subsets, which is given by a {\it Stirling number of the second kind}. 
That is
\begin{equation}
N(n_c,m)=\left\{\begin{array}{c}n_c\\m\end{array}\right\},
\label{stirling}
\end{equation}
where we have used the conventional notation. For completeness,
we summarise the properties of these numbers in 
appendix~\ref{app:stirling} (see also e.g.~\cite{mathworld,digilab}).
Rewriting eq.~(\ref{sumzero3}) as
\begin{equation}
\sum_{D'}R_{DD'}=-\sum_{m=1}^n(-1)^m(m-1)!\left\{
\begin{array}{c}n_c\\m\end{array}\right\},
\label{rddzero1}
\end{equation}
this immediately gives zero by a known identity of Stirling numbers 
(eq.~(\ref{stirlingid}) in appendix~\ref{app:stirling}).
This completes the proof of the zero sum row property, eq.~(\ref{zerosum}).

The interpretation of the zero sum row property has been discussed 
in~\cite{Gardi:2010rn}. From eq.~(\ref{Ctilde}), we see that this property
implies a symmetry of the exponent of the Wilson-line correlator under
the transformation
\begin{equation}
C(D')\rightarrow C(D')+K,
\label{coltrans}
\end{equation}
for the conventional colour factor of all the diagrams $D'$ in a given web, where $K$ is
a constant independent of which diagram one is considering. 
In other words, the part of each colour factor which does not depend
on the ordering of gluon attachments -- thus contributing equally to all $C(D')$ in the web  --   does not enter the exponent. These symmetric terms are instead generated by the explicit exponentiation of lower order webs. 

In fact, we can go further than the result of eq.~(\ref{zerosum}), in specific cases
in which webs contain subsets of diagrams of differing degree of planarity.
This is the subject of the following section.

%%%%%%%%%%%%%%%%%%%%%%%%%%%%%%%%%%%%%%%%
\section{Constraints from the planar limit}
\label{sec:planar}

\begin{figure}[htb]
\begin{center}
\vspace*{10pt}
\scalebox{1.0}{\includegraphics{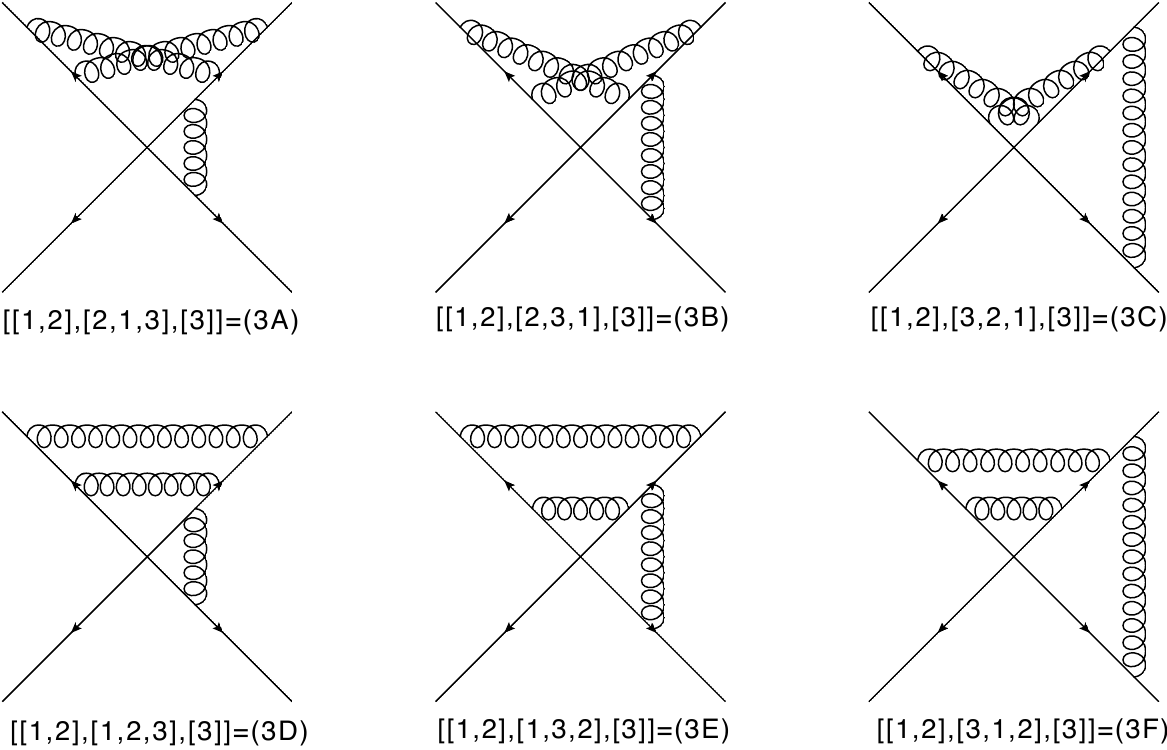}}
\caption{Example diagrams in which three parton lines are linked by three
 parton emissions.}
\label{3lsix}
\end{center}
\end{figure}

In this section, we consider specific cases in which the zero sum row
property of eq.~(\ref{zerosum}) can be specialized. That is, it is sometimes
possible to prove a stronger statement, namely that the zero sum property holds also for
\begin{equation}
\sum_{D'\in{\cal D}}R_{DD'}=0,
\label{zerosumsp}
\end{equation}
where ${\cal D}$ is a subset of diagrams in the web. To illustrate this,
we use the example web shown in figure~\ref{3lsix} (also considered 
in~\cite{Gardi:2010rn}, where it was used to illustrate the cancellation of subdivergences). The corresponding web mixing matrix is given by
\begin{equation}
\frac{1}{6}\left(\begin{array}{rrrrrr}3&0&-3&-2&-2&4\\-3&6&-3&1&-2&1
\\-3&0&3&4&-2&-2\\0&0&0&1&-2&1\\0&0&0&-2&4&-2\\0&0&0&1&-2&1\end{array}\right)
\left(\begin{array}{c}C(3A)\\C(3B)\\
C(3C)\\C(3D)\\C(3E)\\C(3F)\end{array}\right),
\label{1-6mat}
\end{equation}
where we include the vector of conventional colour factors so as to make
clear the ordering of the matrix. As can be seen from eq.~(\ref{1-6mat}),
the subsets of diagrams (3A)-(3C) and (3D)-(3F) each separately 
satisfies the condition of eq.~(\ref{zerosumsp}). For example, in the second 
row of the matrix, the former and latter three entries give $-3+6-3=0$
and $1-2+1=0$ respectively. This hints at an extra structure of web mixing 
matrices over and above the zero sum row property of eq.~(\ref{zerosum}), and
in fact this extra structure can be understood, and shown to hold in general, by appealing to the planar limit of non-Abelian gauge theory, as we now show.

\begin{figure}[htb]
\begin{center}
\vspace*{20pt}
\scalebox{1.0}{\includegraphics{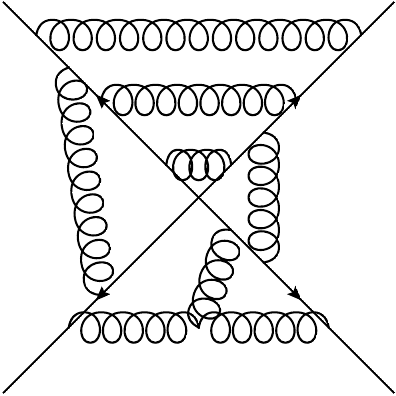}}
\caption{Example diagram which survives in the planar limit. Gluons can only connect adjacent parton legs.}
\label{planex}
\end{center}
\end{figure}

We consider the limit in which the number of colours $N_c$ becomes large, with
the \hbox{\it 't Hooft coupling} $g_s^2N_c$ held fixed. As is well 
known~\cite{'tHooft:1973jz}, only planar diagrams contribute in this limit. 
The structure of those diagrams which contribute to the exponent of the 
soft-gluon amplitude are discussed, for example, 
in~\cite{Bern:2005iz}. Considering planar diagrams, soft gluons may only connect 
adjacent parton legs, an example of which can be seen in figure~\ref{planex}.
Thus, each diagram corresponds to a set of wedges. As discussed 
in~\cite{Bern:2005iz}, each wedge generates half the infrared singularities 
of a Sudakov form factor associated with the partons $i$, $i+1$ forming 
the wedge. It may also be shown to all orders that the soft gluon 
amplitude becomes proportional to the identity matrix $\delta_{IJ}$ 
in the space of possible colour flows. That is, although there is more than 
one possible colour flow possible in any given multiparton diagram, the 
different possible colour flows do not interfere with each other at leading 
order in the large-$N_c$ expansion. As a consequence, the colour factors 
$C(G)$ for gluon subdiagrams commute with each other. Furthermore, $n$-parton 
scattering in the planar limit becomes a set of $n$ copies of the two-eikonal 
line case, with each wedge in the diagram behaving as two Wilson lines 
joined by a colour singlet cusp. 

\begin{figure}
\begin{center}
\scalebox{1.0}{\includegraphics{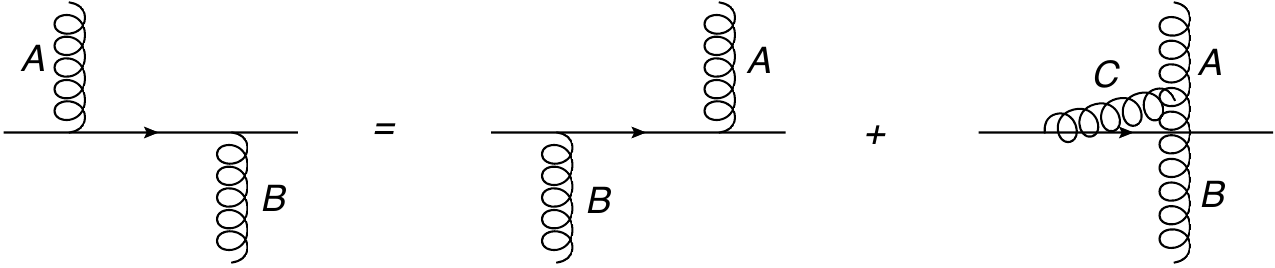}}
\caption{Graphical representation of the commutation property of gluons on 
parton line $i$ in the planar limit. Gluons above and below the line 
represent connections to lines $i+1$ and $i-1$. The second term on the 
right-hand side is non-planar and thus can be neglected.}
\label{comid}
\end{center}
\end{figure}

\begin{figure}
\begin{center}
\scalebox{1.0}{\includegraphics{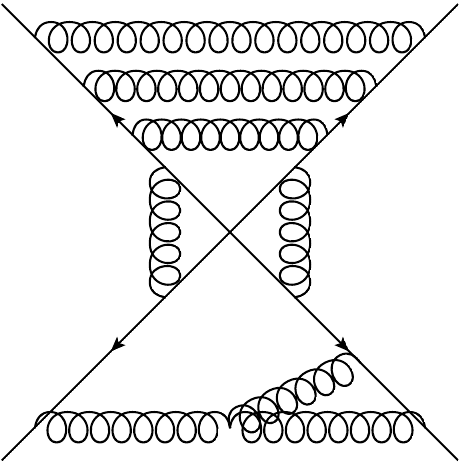}}
\caption{Diagram whose colour factor is equivalent to that of 
figure~\ref{planex} in the planar limit, but where the subdiagrams in 
each wedge are mutually decoupled.}
\label{planex2}
\end{center}
\end{figure}

How does the above structure emerge from the methods used 
in~\cite{Gardi:2010rn} and in the present paper?
Firstly, one notes that on any given parton line $i$, there are gluons which 
connect this line to the adjacent lines $i+1$ and $i-1$ (gluons that connect to non-adjacent lines would form non-planar diagrams, and thus should not be considered). We may draw these pictorially as 
gluons lying on either side of the line $i$, as in figure~\ref{comid}. 
One may now commute all lower gluons to the left of the upper gluons. Such 
a commutation is shown pictorially in figure~\ref{comid}, and corresponds 
to writing
\begin{equation}
T_i^AT_i^B=T_i^BT_i^A+[T_i^A,T_i^B],
\label{comideq}
\end{equation}
where $T_i^A$ and $T_i^B$ are the colour matrices for the upper and lower 
gluons respectively. Using the Lie algebra definition
\begin{equation}
[T_i^A,T_i^B]=if^{ABC}T_i^C,
\label{Lie}
\end{equation}
the second term on the 
right-hand-side of eq.~(\ref{comideq}) is equivalent to a three-gluon coupling, which links all 
three lines $i$, $i-1$ and $i+1$. Such a contribution is necessarily non-planar, and 
thus can be neglected in the large-$N_c$ limit. It follows that the colour 
factor of any diagram of the form of figure~\ref{planex} is the same as that 
for an equivalent diagram in which the gluon subdiagrams in each wedge are 
decoupled from each other, as shown for example in figure~\ref{planex2}.

As already discussed above, in the planar limit the colour factors for 
individual subdiagrams $C(G)$ are proportional to the identity matrix in 
colour space, thus commute with each other. The ECF of 
any subdiagram can then be determined using the equations~(\ref{modcol3}) 
or~(\ref{gatheral}) (rather than~(\ref{modcol2}) or~(\ref{colfacfin2})). 
It then follows from eq.~(\ref{modcol3}) and the above discussion that the 
ECF of a given planar diagram is equal to the ECF of an equivalent diagram in which the subdiagrams in each wedge 
are decoupled from each other (for example the ECF's of 
figures~\ref{planex} and~\ref{planex2} are the same, as well as their 
conventional colour factors). To see this from eq.~(\ref{modcol3}), one 
must replace all colour factors $C(G)$ on the right-hand-side with the 
colour factors of their equivalent diagrams in which the subdiagrams are 
decoupled. The left-hand side then represents the ECF 
of the decoupled graph corresponding to $G$. 

Next, we shall use eq.~(\ref{gatheral}) to establish the following lemma: \emph{in the planar limit the ECF's for reducible graphs are zero}. 
By reducible graphs we mean diagrams such as those in figure~\ref{17-20}, whose colour factors can be decomposed as 
\begin{equation}
C(G)=C(G_1)C(G_2)\,,
\label{CGdecomp}
\end{equation}
where $G_1$ and $G_2$ may themselves be further reducible. 
The notion of irreducibility of webs is well-known in the case of two-eikonal lines, and in the planar limit this notion extends to $n$-parton scattering. The proof of this result is 
essentially the same as that given in e.g.~\cite{Gatheral:1983cz} for 
the two-line case, but simplified slightly due to eq.~(\ref{NDres}),
which was not recognized in~\cite{Gatheral:1983cz}.

One proceeds by induction, after noting that the decompositions of a given 
diagram $G$ may be separated into the {\it trivial decomposition}  -- the one 
containing $G$ itself -- and {\it proper decompositions} in which $G$ 
genuinely reduces into lower order diagrams. One may then rewrite 
eq.~(\ref{gatheral}) as
\begin{equation}
  C(G) = \widetilde{C}(G) + \sum_{ \{ m'_H \} } 
  \left( \prod_H \widetilde{C}(H)^{\, m'_H} \right) \, ,
\label{colfacts2}
\end{equation}
where the prime denotes proper decompositions. The inductive hypothesis 
assumes that the vanishing of ECF's for reducible 
diagrams has already been shown up to some order, so that each of the 
factors $\widetilde{C}(H)$ on the right-hand side of eq.~(\ref{colfacts2}) 
corresponds to an irreducible diagram. One may then show that $\widetilde{C}(G)$ 
on the left-hand side is zero, if $G$ is reducible. 

\begin{figure}
\begin{center}
\scalebox{1.0}{\includegraphics{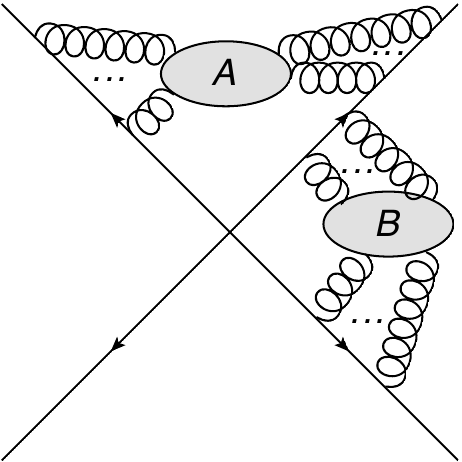}}
\caption{General form of a reducible diagram, involving subdiagrams connected 
to the same parton line.}
\label{reducible}
\end{center}
\end{figure}

A general form of a reducible diagram involving subdiagrams connected to the 
same parton line is shown in figure~\ref{reducible}. More generally, one 
may have a further subdiagram connecting the third and fourth lines in 
figure~\ref{reducible}, although the following argument is easily generalized 
so that it is sufficient to consider the present form only.
For such a diagram one may write
\begin{equation}
C(G)=C(AB)=C(A)C(B),
\label{CGdec}
\end{equation}
so that eq.~(\ref{colfacts2}) becomes 
\begin{equation}
  C(A)C(B) = \widetilde{C}(AB) + \sum_{ \{ m'_H \} } 
  \left( \prod_H  \, \widetilde{C}(H)^{\, m'_H} \right) \, .
\label{multAB}
\end{equation}
Each subdiagram $H$ on the right-hand side is irreducible, thus contributes 
to $A$ or $B$ but not both (otherwise this would contradict the 
reducibility of $G$). One may thus decompose the product over decompositions
into two separate products (one for each subdiagram):
\begin{align}
C(A)C(B) &= \widetilde{C}(AB) +\left[ \sum_{ \{ m_H^A \} } 
  \left( \prod_H  \, \widetilde{C}(H)^{\, m_H^A} 
  \right) \right] \left[ \sum_{ \{ m_H^B \} } 
   \left( \prod_H 
  \, \widetilde{C}(H)^{\, m_H^B} \right) \right],
\label{multAB3}
\end{align}
where we have separated the sum over proper decompositions into separate 
sums for each subdiagram. By eq.~(\ref{gatheral}), the second term on the 
right-hand side of eq.~(\ref{multAB3}) is clearly $C(A)C(B)$, so that 
\begin{equation}
\widetilde{C}(AB)=0.
\label{CAB0}
\end{equation}
This result shows that the ECF's for reducible graphs 
at ${\cal O}(\alpha_S^{n+1})$ are zero provided this is true 
at ${\cal O}(\alpha_S^n)$. Given that diagrams are irreducible at one 
loop order, this proves the result that the ECF's for 
reducible graphs vanish in the planar limit to all orders. 

Given that we have also argued above that to leading order in the large-$N_c$ limit, the ECF of a 
general planar diagram is the same as that of an equivalent diagram in 
which subdiagrams in different wedges are decoupled from each other -- a reducible diagram -- it follows that the ECF for any diagram which contains 
emissions in more than one wedge is zero in the large-$N_c$ limit. 
Thus, in this limit the exponent of any Wilson-line correlator contains only diagrams in which emissions occur in a single wedge, and where the subdiagrams are two-eikonal line irreducible with respect to the Wilson lines forming the wedge. 
This agrees with the argument of~\cite{Bern:2005iz}, that in the 
planar limit, the $n$ eikonal line amplitude is given by $n$ copies of the two eikonal line amplitude. 

Returning to the example of figure~\ref{3lsix}, the above discussion tells
us that to leading order in the large-$N_c$ limit, $\widetilde{C}(D)=0$ for all diagrams $D$ in the web, as they all involve
emissions in more than one wedge. However, we may also note that diagrams
(3A)-(3C) are non-planar, whilst diagrams (3D)-(3F) are planar. Thus, in the
large $N_c$ limit, one has $C(3A)=C(3B)=C(3C)=0$, and the commuting property
of gluons in the planar limit implies $C(3D)=C(3E)=C(3F)$ (where each is
non-zero, as these are planar diagrams). 
We may write eq.~(\ref{Ctilde}) in the large $N_c$ limit as
\begin{equation}
0=\sum_{D'}R_{DD'}C(D')=\sum_{D'\in{\cal D}_{1}}R_{DD'}C(D')+
\sum_{D'\in{\cal D}_{2}}R_{DD'}C(D'),
\label{CtildelargeN}
\end{equation}
where ${\cal D}_1=\{(3A),(3B),(3C)\}$ and ${\cal D}_2=\{(3D),(3E),(3F)\}$.
The first term on the right-hand side is zero using the above results for
$C(3A)$ etc., and the second term can be simplified using $C(D')=C(3D)$ for each term (up to corrections that are subleading in $N_c$) to give
\begin{equation}
C(3D)\sum_{D'\in{\cal D}_2}R_{DD'}=0,
\label{CtildelargeN2}
\end{equation}
and therefore
\begin{equation}
\sum_{D'\in{\cal D}_2}R_{DD'}=0.
\label{CtildelargeN3}
\end{equation}
This has the form of eq.~(\ref{zerosumsp}), where the sum is over the subset 
of diagrams in the web which are planar. Note that eq.~(\ref{CtildelargeN3})
applies to any row of the mixing matrix. Furthermore, the full zero sum row
property of eq.~(\ref{zerosum}) then also implies
\begin{equation}
\sum_{D'\in{\cal D}_1}R_{DD'}=0.
\label{CtildelargeN4}
\end{equation}
This is trivially satisfied for the latter three rows of the mixing matrix
(eq.~(\ref{1-6mat})), in which the first three elements are zero (as 
essentially follows from the fact that one cannot make a diagram with
a crossed gluon pair out of partitions of a diagram in which gluons are
not crossed). However, eqs.~(\ref{CtildelargeN3}) and~(\ref{CtildelargeN4}) 
give us useful information in the first three rows of the mixing matrix, namely
that the first and last three entries of each row must separately sum to zero.

The generalization of the above remarks is straightforward, and can be compactly 
stated as follows: \emph{in any multiparton web -- closed set of diagrams related by permutations -- the zero sum property holds separately for the planar and nonplanar diagrams in each row.}
This is a stronger condition than the overall zero sum row property, and provides an
example of the rich substructure of the web mixing matrices. Clearly, these matrices must have further interesting substructure related to the cancellation of subdivergences~\cite{Gardi:2010rn}, which remains to be fully understood. 

Note that in section~\ref{sec:zerosum} we used pure mathematical arguments
to arrive at the zero sum row property -- this was an exercise in enumerative
combinatorics. In contrast, in the present section we have applied a known piece of physics -- colour structures in the large-$N_c$ limit -- to pin down additional properties. 
This suggests two main avenues for further research into the properties of webs. Either one may adopt a purely mathematical approach, based on the combinatoric result of eq.~(\ref{RDD}). Or, one may continue to investigate known examples of physical behaviour, and translate these into constraints on the mixing matrix. 

%%%%%%%%%%%%%%%%%%%%%%%%%%%%%%%%%%%%%%%
\section{Discussion}
\label{sec:conclude}

In this paper we have studied the exponents of Wilson-line correlators, following
on from the results of~\cite{Gardi:2010rn,Mitov:2010rp}, which generalized the
concept of webs from two-parton to multiparton scattering. 
In particular, we have examined the properties of web mixing matrices, 
whose existence was derived in~\cite{Gardi:2010rn}, and proved in full generality the properties of \emph{idempotence} and \emph{zero sum rows} that were conjectured there. This is an important step in establishing the properties of webs, and thus in understanding the structure of exponentiation in non-Abelian gauge theory amplitudes. 

The importance of the web mixing matrices stems from the fact that they encapsulate the correlation between colour and kinematic dependence in the exponent, correlation which becomes highly non-trivial for multiparton amplitudes (or Wilson-line correlators) at general $N_c$. As emphasized 
in~\cite{Gardi:2010rn} these matrices are responsible, in particular, for intricate cancellations of subdivergences rendering the singularity structure of webs consistent with that required by the renormalization properties of the multi-eikonal vertex~\cite{Gardi:2010rn,Mitov:2010rp}. 

The proof of idempotence presented here relies on the replica trick formalism. This formalism was already used in~\cite{Gardi:2010rn} to show the existence of web mixing matrices and to compute them. 
Here, it allows the idempotence property to be derived in an elegant fashion, by replicating the theory twice and then showing that the doubly-replicated theory is essentially equivalent to the singly-replicated one. In this analysis idempotence of mixing matrices ultimately derives from the idempotence of the replica-ordering operator ${\cal R}$.

The physical interpretation of the idempotence property~\cite{Gardi:2010rn} is that the web mixing matrices act as \emph{projection operators}. Their space of eigenvalues is composed exclusively of $0$ and~$1$, both of which are generically degenerate (note that there is always at least one zero eigenvalue, as a consequence of the zero sum row property). The mixing matrices thus select those linear combinations of kinematic and colour factors which correspond to the eigenvalue~$1$ to enter the exponent, while removing all those which correspond to eigenvalue~$0$. The latter are precisely the terms that are generated by expanding the exponential containing lower order webs. 
Ref.~\cite{Gardi:2010rn} demonstrated (using non-trivial three-loop examples) that the required cancellations of subdivergences in the exponent indeed take place in the particular linear combinations of kinematic functions of eigenvalue~$1$. The general structure responsible for this cancellation remains to be fully explored.

Another important result of the present paper is the explicit combinatorial formula for the mixing matrices in eq.~(\ref{RDD}) in terms of \emph{overlap functions}. An overlap function $\ip{D'}{P_D}$ counts the number of ways a given diagram $D'$ in a closed set of diagrams (web) can be made out of a particular decomposition $P_D$ of another diagram ($D$) in the set.
This explicit formula would be a convenient starting point for further mathematical exploration of webs.
 
Indeed, starting with this formula, the zero sum rows property was shown in section \ref{sec:zerosum}  to correspond to known results from enumerative combinatorics involving Stirling numbers of the second kind.
This analysis strongly suggests that further results from enumerative combinatorics, and in particular from the theory of integer partitions, will prove useful in studying the properties of webs. This is an interesting connection, which we are continuing to investigate. 

We have further used the zero sum rows property, in conjunction with 't Hooft's large-$N_c$ limit, to derive a stronger result, namely that in any multiparton web containing both planar and non-planar diagrams, the zero sum property holds, in each row, separately for elements of the mixing matrix corresponding to the planar and the non-planar diagrams. This result relies on the fact that in the planar limit multi-parton webs reduce to a simple sum over two-parton webs between adjacent partons (wedge), excluding from the exponent any diagram that incorporates exchanges in more than one wedge. Despite the trivial nature of webs in this limit, it provides a useful  constraint on web mixing matrices valid for general~$N_c$. This is another promising avenue for further exploration of webs. 

Finally, it is clear from the analysis of Ref.~\cite{Gardi:2010rn}, as well as section \ref{sec:planar} here, that web mixing matrices must have a rather subtle substructure, which remains to be fully explored. Formulating the properties of webs would very likely be a necessary prerequisite for determining the all-order structure of multiparton scattering amplitudes. 
In particular, these would be essential for a detailed understanding of the singularity structure of webs, and how this conforms with the renormalization properties of the multi-eikonal vertex, which involves nested commutator structures of lower-order counterterms and webs~\cite{Mitov:2010rp}. 
Research in this direction is ongoing. 

\vspace*{30pt}

\acknowledgments

We thank Eric Laenen for useful discussions, and Lorenzo Magnea for comments
on the manuscript. 
CDW is supported by the STFC postdoctoral fellowship ``Collider Physics at 
the LHC''. We have used JaxoDraw~\cite{Binosi:2008ig,Binosi:2003yf} 
throughout the paper.

\vspace*{30pt}

\appendix

%%%%%%%%%%%%%%%%%%%%%%%%%%%%%%%%%%%%%%%%%%
\section{Inverted ECF formula in Refs.~\cite{Gardi:2010rn} and~\cite{Mitov:2010rp}\label{sec:inverse_formulae_comparison}}

The purpose of this section is to elucidate the relation between our approach to webs, as first described in~\cite{Gardi:2010rn}, and the one of reference~\cite{Mitov:2010rp}. These two papers, which appeared simultaneously, have both presented formulae which express
the conventional colour factors in terms of the ECF's. While using somewhat different considerations and different notation, the two have the same content as we now explain. 

While the primary approach of ref.~\cite{Gardi:2010rn}, much like the present paper, has been based on the replica trick, yielding an explicit formula for ECF's in terms of the conventional ones, 
Section 4.2 in~\cite{Gardi:2010rn} presented an inverse relation, expressing
the conventional colour factors in terms of the ECF's, which we shall now recall.
   
To this end one may first consider the set $\{H\}$ of all possible subdiagrams at all orders in perturbation theory. Each decomposition $P$ of a general graph $D$ can be uniquely labelled by a set of numbers $m_H$, each representing how many times subdiagram $H$ occurs as an element of the 
decomposition ($m_H=0$ if $H$ does not occur). The conventional colour factors are then related to the 
ECF's by the following formula, as given in~\cite{Gardi:2010rn}:
\begin{equation}
C(D)=\sum_{\{m_H\}}\frac{N_{D|\{m_H\}}}{n!}\left(\prod_Hm_H!\right)^{-1}
\left[\widetilde{C}(H_1)^{m_1}\widetilde{C}(H_2)^{m_2}\ldots +\text{perms}
\right],
\label{colfacfin}
\end{equation}
where $n=\sum_H m_H$ and $N_{D|\{m_H\}}$ represents the number of ways
in which diagram $D$ can be formed from the decomposition specified by
$\{m_H\}$. This is taken to be zero for those decompositions which
cannot lead to $D$; otherwise this multiplicity factor is given by
\begin{equation}
N_{D|\{m_H\}}=\prod_Hm_H!\,.
\label{NDres}
\end{equation}
One may form $D$ in multiple ways by permuting identical subdiagrams,
and there are $m_H!$ such permutations for each subdiagram $H$, so that 
eq.~(\ref{colfacfin}) may be simplified to
\begin{equation}
C(D)=\sum_{\{m_H\}}\frac{1}{n!}
\left[\widetilde{C}(H_1)^{m_1}\widetilde{C}(H_2)^{m_2}\ldots +\text{perms}
\right].
\label{colfacfin2}
\end{equation}
An identical result was also given in~\cite{Mitov:2010rp}, and it is our purpose here to convert the results of that paper into the present notation, so as to demonstrate this equivalence. 

Reference~\cite{Mitov:2010rp} considers the correlator of a
number of Wilson lines meeting at a common vertex, writing it as follows:
\begin{equation}
A[C_i]=\exp\left(\sum_{i=1}^\infty w^{(i)}\right),
\label{Aexp}
\end{equation}
where $w^{(n)}$ collects all diagrams at ${\cal O}(\alpha_s^n)$ in the
exponent, and implicitly contains a factor $\alpha_s^n$. This is further
decomposed as
\begin{equation}
w^{(i)}=\sum_Ew_E^{(i)},
\end{equation}
where the sum is over sets of numbers $E=\{e_1\ldots e_L\}$, such that 
$e_i$ is the number of gluon attachments on parton line $i$. The sum over
$E$ is thus equivalent to summing over closed sets of diagrams related
by gluon permutations, as discussed in the previous section. Each set $E$
corresponds to a distinct web according to our definition. In position 
space, each web may be written as~\cite{Mitov:2010rp}
\begin{equation}
w_E^{(i)}={\cal I}_E[{\cal W}_E^{(i)}],
\label{wE}
\end{equation}
where ${\cal I}_E$ denotes the integrals over the positions of the gluon
emissions on each parton line (with appropriate measure and limits), and
${\cal W}_E^{(i)}$ the integrand, containing both kinematic and colour
information. With this notation,~\cite{Mitov:2010rp}
gives the following formula for the exponent of the soft gluon amplitude
at ${\cal O}(\alpha_s^{N+1})$:
\begin{equation}
w^{(N+1)}=\sum_E\sum_{D_E^{(N+1)}}\left\{D_E^{(N+1)}-{\cal I}_E\left[\sum_{m=2}^{N+1}
\sum_{\Omega_m(D_E^{(N+1)})}\frac{\prod_c m_c!}{m!}\sum_{\text{sym}}
{\cal W}_{E_m}^{(i_m)}\ldots{\cal W}_{E_1}^{(i_1)}\right]\right\}.
\label{eq17}
\end{equation}
This is an iterative formula, which explicitly relates the exponent at 
a given order to lower order webs. Some explanatory comments are in order. 
The first sum on the right-hand side is over sets of numbers $E$ as
explained above i.e. a sum over distinct closed sets of diagrams related
by gluon permutations. The second sum is over all diagrams $D_E^{N+1}$ 
which have gluon attachments characterized by $E$, and which are
${\cal O}(\alpha_s^{N+1})$. That is, the sum goes over all diagrams
within the closed set labelled by $E$. In what follows, we will shorten
this notation and simply write $D\equiv D_E^{(N+1)}$. The first term in 
the brackets is then the complete expression for the diagram $D$, which in
our notation is given by
\begin{equation}
D\equiv{\cal F}(D)C(D),
\label{Dres}
\end{equation}
where as usual ${\cal F}(D)$ and $C(D)$ are the kinematic and (conventional)
colour parts respectively. In the second term in the curly brackets, 
${\cal I}_E$ represents the position space integrals over the eikonal 
attachments corresponding to the particular diagram $D$. The integrand
is in the square brackets, and consists of a sum over products of lower 
order webs. The index $m$ labels the number of lower order webs in each 
term, and ${\cal W}_{E_i}^{(i_m)}$ is the integrand of a lower order 
function $w^{(i)}$, including both colour and kinematic information. The 
index $\Omega_m(D_E^{(N+1)})$ runs over over all sets of lower order 
web integrands, whose combination gives diagrams which are 
topologically equivalent to $D$. There is a further combinatoric factor 
for each number of webs $m$, where each lower order web integrand is 
assumed to occur $m_c$ times. Finally, there is a sum over all 
distinguishable permutations of the ${\cal W}$ factors, denoted
by $\sum_{\text{sym}}$. 

It is straightforward to show that eq.~(\ref{eq17}) is equivalent to
eq.~(\ref{colfacfin2}). Firstly, we may write the sum over 
distinguishable permutations as a sum over all permutations of
the lower order web factors:
\begin{equation}
\sum_{\text{sym}}{\cal W}_{E_m}^{(i_m)}\ldots{\cal W}_{E_1}^{(i_1)}=
\left(\prod_c\frac{1}{m_c!}\right)
\left(\left[{\cal W}_{E_1}^{(i_1)}\right]^{m_1}\ldots
\left[{\cal W}_{E_n}^{(i_n)}\right]^{m_n} +{\rm perms.}\right),
\label{perms}
\end{equation}
where we have counted all permutations (including indistinguishable ones) on
the right-hand side, and the inverse factorial factors correct for the fact that we 
have overcounted permutations which are related by interchanging 
identical web factors. Also, $n$ is the number of distinct
lower order webs. We may also write
\begin{equation}
{\cal I}_E\,\left[{\cal W}_{E_1}^{(i_1)}\right]^{m_1}\ldots
\left[{\cal W}_{E_n}^{(i_n)}\right]^{m_n}={\cal F}_D
\left[\widetilde{C}(H_1)^{m_1}\ldots\widetilde{C}(H_n)^{m_n}\right].
\label{Hdef}
\end{equation}
That is, the integral over the kinematic part of the product of 
${\cal W}$ factors is, by construction, the kinematic part of $D$. 
The colour part is given by the product of the colour factors of the 
individual webs: these are exponentiated colour factors rather than conventional ones. We have denoted these lower order web 
diagrams by $H_i$ on the right-hand side of eq.~(\ref{Hdef}), 
where each diagram occurs $m_H$ times with $\sum_{m_H}=m$. 
We may then recognise the sum in the square brackets in eq.~(\ref{eq17}) 
as a sum over all possible decompositions $\{m_H\}$, as defined above, 
but where these have at least two elements. 
Denoting such {\it proper decompositions} by $\{m'_H\}$, the contents of the curly bracket in eq.~(\ref{eq17}) 
(which, by definition, is $\widetilde{C}(D){\cal F}(D)$) thus has the form
\begin{equation}
\widetilde{C}(D){\cal F}(D)=C(D){\cal F}(D)-\sum_{\{m'_H\}}\frac{1}{m!}
{\cal F}(D)\left[\widetilde{C}(H_1)^{m_1}\ldots\widetilde{C}(H_n)^{m_n}
+\text{perms.}\right],
\label{eq17b}
\end{equation}
which gives
\begin{equation}
\widetilde{C}(D)=C(D)-\sum_{\{m'_H\}}\frac{1}{m!}
\left[\widetilde{C}(H_1)^{m_1}\ldots\widetilde{C}(H_n)^{m_n}
+\text{perms.}\right].
\label{modcolsterman}
\end{equation}
Note that the left-hand side corresponds to the product of exponentiated 
colour factors in the {\it trivial decomposition} of $D$, consisting only 
of $D$ itself. We may thus combine this with the sum over proper 
decompositions to get
\begin{equation}
C(D)=\sum_{\{m_H\}}\frac{1}{m!}
\left[\widetilde{C}(H_1)^{m_1}\ldots\widetilde{C}(H_n)^{m_n}+\text{perms.}\right],
\label{modcol2sterman}
\end{equation}
which is eq.~(\ref{colfacfin2}) (after relabelling $m$ to $n$). 

It is also useful (see e.g. section~\ref{sec:planar})  to determine the specific form of
eqs.~(\ref{modcol2}) and~(\ref{colfacfin2}), in cases where the
conventional colour factors of distinct subdiagrams commute with each 
other. In eq.~(\ref{modcol2}) this allows one to write
\begin{equation}
\sum_\pi C(g_{\pi_1})\ldots C(g_{\pi_{n(P)}})=n!\prod_{g\in P}C(g)
\label{perms2}
\end{equation}
where each permutation has been rearranged to give the same ordering, and
there are $n!$ such permutations, so that eq.~(\ref{modcol2}) becomes
\begin{equation}
\widetilde{C}(G)=\sum_P(-1)^{n(P)-1}(n(P)-1)!\prod_{g\in P} C(g),
\label{modcol3}
\end{equation}
where the product is over all subdiagrams in the decomposition $P$. Also for the inverse relation (\ref{colfacfin2}) a similar simplification occurs in the commuting case: one has 
\begin{equation}
\widetilde{C}(H_1)^{m_1}\widetilde{C}(H_2)^{m_2}\ldots \widetilde{C}(H_n)^{m_n}+\text{perms}=n!\,\prod_H
\widetilde{C}(H)^{m_H},
\label{colfaccom}
\end{equation}
so that eq.~(\ref{colfacfin2}) becomes
\begin{equation}
C(G)=\sum_{\{m_H\}}\prod_H\widetilde{C}(H)^{m_H}.
\label{gatheral}
\end{equation}

%%%%%%%%%%%%%%%%%%%%%%%%%%%%%%%%%%%%%%%%%%
\section{Stirling numbers of the second kind}
\label{app:stirling}

In section~\ref{sec:zerosum}, we use Stirling numbers of the second kind in the proof of
the zero sum row property of web mixing matrices. As these numbers may not be widely 
familiar to all readers, we provide a short summary of their properties in this appendix (see also e.g.~\cite{mathworld,digilab}).
\begin{table}[htb]
\begin{center}
\begin{tabular}{cc|cccccc}
%\hline
&&$m$&$\rightarrow$&   & &  \\
&&1 &2 & 3 & 4& 5 &6\\
\hline
$n$&1& 1 & & & &  \\
$\downarrow$&2& 1 & 1 & & & &  \\
&3& 1 & 3 & 1 & & & \\
&4& 1 & 7 & 6 & 1 & & \\
&5& 1 & 15 & 25 & 10 & 1 & \\
&6& 1 & 31 & 90 & 65 & 15 & 1\\
\end{tabular}
\caption{The first few Stirling numbers of the second kind.}
\label{stirlings}
\end{center}
\end{table}

The Stirling number of the second kind, conventionally written 
${\small \left\{\begin{array}{c}n\\m\end{array}\right\}}$, counts
the number of partitions of a set of $n$ objects into $m$ non-empty subsets.
Based on this definition it is straightforward to construct a recursion relation: 
one can count the number of partitions of $n$ objects into $m$ (non-empty) subsets as follows: separate one object from the other $n-1$. This object can either be added to one of $m$ partitions of the other $n-1$ objects -- this yields $m{\small \left\{\begin{array}{c}n-1\\m\end{array}\right\}}$ distinct possibilities -- or form a (single-object) partition by itself -- yielding ${\small \left\{\begin{array}{c}n-1\\m-1\end{array}\right\}}$ additional possibilities.
It therefore follows that
\begin{equation}
\left\{\begin{array}{c}n\\m\end{array}\right\}=
\left\{\begin{array}{c}n-1\\m-1\end{array}\right\}+
m\left\{\begin{array}{c}n-1\\m\end{array}\right\}\,.
\label{recur}
\end{equation}
The solution of this recursion can be written in a closed form:
\begin{equation}
\left\{\begin{array}{c}n\\m\end{array}\right\}=\frac{1}{m!}
\sum_{j=0}^m(-1)^j\left(\begin{array}{c}m\\j\end{array}\right)
(m-j)^n\,.
\label{recursol}
\end{equation}
The first few values are given in table~\ref{stirlings}; one may readily verify that they admit the above relations.

A useful additional identity, which is being used here in the proof of the 
zero sum row property, is the following (see e.g.~\cite{mathworld}):

\begin{equation}
\sum_{m=1}^n(-1)^m(m-1)!\left\{\begin{array}{c}n\\m\end{array}\right\}=0.
\label{stirlingid}
\end{equation}

\bibliographystyle{JHEP}
\bibliography{refs1}
\end{document}